\documentclass[12pt,preprint]{aastex}

\def\mum{${\rm \mu m}$}
\def\sta{C$_{84}$H$_{24}$}
\def\stb{C$_{90}$H$_{30}$}
\def\stc{C$_{102}$H$_{26}$\,C$_{2v}$}
\def\std{C$_{102}$H$_{26}$\,C$_{2h}$}
\def\ste{C$_{110}$H$_{30}$}
\def\stf{C$_{120}$H$_{36}$}
\def\stref{C$_{96}$H$_{24}$}
\newcommand{\HII}{\ion{H}{2}}

\shorttitle{Large irregular PAHs.}
\shortauthors{Bauschlicher et al.}
\begin{document}
\title { The Infrared Spectra of Very Large                              
Irregular polycyclic aromatic hydrocarbons (PAHs): \\
Observational Probes of Astronomical PAH Geometry, Size and Charge}
\author{Charles W. Bauschlicher, Jr.\altaffilmark{1}, 
Els Peeters\altaffilmark{2,3}, Louis J. Allamandola\altaffilmark{4}}

\altaffiltext{1}{NASA-Ames Research Center, Space Technology Division,
Mail Stop 230-3, Moffet Field, CA 94035, USA;
Charles.W.Bauschlicher@nasa.gov}

\altaffiltext{2}{Department of Physics and Astronomy, 
University of Western Ontario, London, ON N6A 3K7, Canada;
epeeters@uwo.ca}

\altaffiltext{3}{SETI Institute, 515 N. Whisman Road, Mountain View,
CA 94043, USA}

\altaffiltext{4}{NASA-Ames Research Center, Space Science Division,
Mail Stop 245-6, Moffet Field, CA 94035, USA;
Louis.J.Allamandola@nasa.gov}

\begin{abstract}
The mid-IR spectra of six large, irregular PAHs with formulae (\sta\, -- \stf) have been computed using Density Functional  Theory (DFT).  Trends in the dominant band positions and intensities are compared to those of large, compact PAHs as a function of geometry, size and charge. Irregular edge moieties that are common in terrestrial PAHs, such as bay regions and rings with quartet hydrogens, are shown to be uncommon in astronomical PAHs. As for all PAHs comprised solely of C and H reported to date, mid-IR emission from irregular PAHs fails to produce a strong CC$_{str}$ band at 6.2 \mum, the position characteristic of the important, class A astronomical PAH spectra. Earlier studies showed inclusion of nitrogen within a PAH shifts this to 6.2 \mum\, for PAH cations. Here we show this band shifts to 6.3 \mum\, in nitrogenated PAH anions, close to the position of the CC stretch in class B astronomical PAH spectra. Thus nitrogenated PAHs may be important in all sources and the peak position of the CC stretch near 6.2 \mum\, appears to directly reflect the PAH cation to anion ratio. 
Large irregular PAHs exhibit features at 7.8 \mum\, but lack them near 8.6 \mum. Hence, the 7.7 \mum\, astronomical feature is produced by a mixture of small and large PAHs while the 8.6 \mum\, band can only be produced by large compact PAHs. As with the CC$_{str}$, the position and profile of these bands reflect the PAH cation to anion ratio. 
\end{abstract}


\section{Introduction}
\label{intro}

Infrared emission from polycyclic aromatic hydrocarbons (PAHs) and closely related species has been detected throughout much of the modern universe \citep[e.g.][]{isoparijs, Helou:normalgal:00, Vermeij:pahs:01,Peeters:colorado, Bregman:05, Brandl:06, Sellgren:07, SmithJD:07, Galliano:08} and detailed PAH models have been developed \citep[e.g.][]{Li:01, Draine:07}.  This emission dominates the mid-IR spectra of many objects \citep[e.g.][]{Uchida:RN:00,Verstraete:prof:01, Peeters:cataloog:02, Peeters:colorado, Onaka:colorado,Brandl:06, Compiegne:07}. The spectral details of this emission vary between different objects and spatially within extended objects \citep[e.g.][and refs. therein]{Peeters:prof6:02, Peeters:colorado, Bregman:05, Rapacioli:05, Berne:07, Sellgren:07}, revealing the nature of the specific PAH molecules present and reflecting the conditions within the emission zones. 

Proper interpretation of these spectra requires a thorough understanding of the spectroscopic properties of PAHs of a size comparable to those which dominate the emission process.  Most previous work on PAH IR spectroscopic properties focused on species containing about 50 or fewer carbon atoms because large PAHs are not readily available for experimental study and computational techniques for such large systems were not practical \citep{Langhoff:neutionanion:96, Bauschlicher:97, Langhoff:substitutedpah:98, Malloci:database, Pathak:08}.  Although large PAH accessibility remains limited, computational capabilities have increased and the spectra of large PAHs can be determined with good precision \citep[e.g.][]{Bauschlicher:C96:02}.  We recently presented and discussed the computational IR spectra of several large, compact, symmetric PAHs ranging in size from C$_{66}$H$_{20}$ to C$_{130}$H$_{28}$ \citep[][hereafter Paper I]{Bauschlicher:vlpahs1}.  These spectra provide new insight into the effect of size and structure on the IR spectroscopic properties for species that are comparable in size to those which are thought to dominate the emitting population of astronomical PAHs.  Here we extend this work to comparably sized PAHs with different structures --in particular, with irregular edge geometries-- and focus on the spectroscopic features which are sensitive to these structural differences.  Since the spectra reflect these structural details in ways that provide deeper understanding of the PAH populations that contribute to the observed astronomical spectra, this information can then be used to gauge the relative importance of PAHs with specific geometrical features in the overall PAH population.
 
This work is presented as follows.  The computational approach and resulting spectra are discussed in Sect.~\ref{model}.  These spectra are applied to astronomical observations in Sect.~\ref{astro}, where specific information about, and constraints on, the charge, size, and structure of the astronomical PAH population are presented.  The paper is concluded in Sect.~\ref{conclusion}.

\clearpage
\begin{figure*}
 \centering \includegraphics[width=\textwidth]{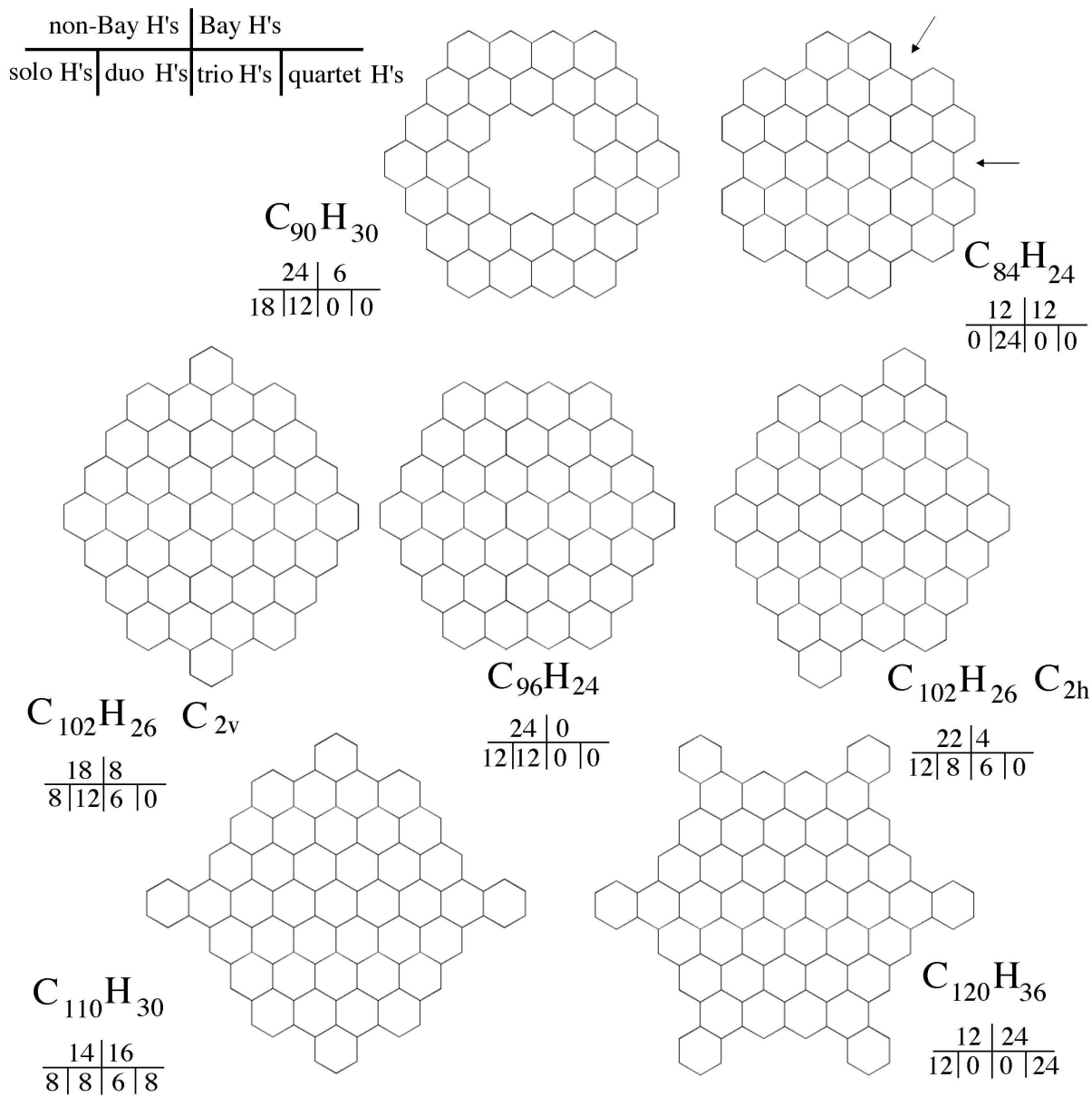}
\caption{The structures of the large irregular PAHs studied in this
paper and the structure of the parent PAH, \stref, from which these
structures are derived. Two of the six bay regions in \sta\, are
indicated by arrows.  The number of H atoms spanning a bay region, and
the number of hydrogen atoms which do not span a bay region, are given
for each PAH.  Likewise, the number of solo, duo, trio, and quartet
hydrogens are also tabulated for each molecule.}
\label{structure}	
\end{figure*}
\clearpage

\section{Computational Methods, Results and Discussion}
\label{model}

Since a detailed description of the computational method is presented in Paper I, the computational approach will just be summarized here.  PAH geometries are optimized and the harmonic frequencies and IR intensities are computed using the B3LYP \citep{Stephens:94} hybrid \citep{Becke:93}  functional in conjunction with the 4-31G basis sets \citep{Frisch:84}. The calculations are performed using the Gaussian 03 computer codes \citep{Frisch:03}.

As in Paper I, attention is focused on the major emission bands.  We do not show all of the data here because the number of bands determined for species of this size is extremely large.  These data are available upon request and will become part of the publicly available Ames PAH IR Spectral Database which is now under construction.  Synthetic spectra are presented in which computed frequencies have been scaled by 0.958, the factor which has been found to bring the computationally determined PAH vibrational frequencies into very good agreement with experimentally measured spectra \citep{Langhoff:neutionanion:96, Bauschlicher:97}.  Intensities are unscaled.  To permit comparison of these absorption spectra with astronomical observations which record emission spectra, the mid-IR bandwidth for a large molecule emitting under interstellar conditions has to be taken into account.  Up to now, this has been taken as about 30 cm$^{-1}$ across the mid-IR.  As discussed in Cami et al. (in preparation), the natural bandwidth can be band dependent.  Here a bandwidth of 30 cm$^{-1}$ is taken for the bands shortward of 9 \mum, and 10 cm$^{-1}$ for the bands longward of 10 \mum, values consistent with the most recent observational and theoretical constraints.  For the 9 to 10 \mum\, region, the FWHM is scaled in a linear fashion (in wavenumber space) from 30 to 10 cm$^{-1}$. Despite these limitations, these idealized spectra can be useful in better understanding the observed interstellar spectra.

\clearpage
\begin{table*}
\small
\caption{\label{t1} The C-H stretching band position maxima
($\lambda$, in $\mu$m), total intensity
(I), intensity per CH$_{bay}$  and intensity per CH$_{non-bay}$ (I(CH)) for the PAHs shown in
Fig. \ref{structure}.  The intensities are in km/mol.}
\begin{center}
\begin{tabular}{l@{\hspace{20pt}}rrrl@{\hspace{20pt}}rrrl@{\hspace{20pt}}rrrr}
\hline \\[-5pt]
Molecule & \multispan4 \hfil Cation \hfil &\multispan4 \hfil  Neutral \hfil &\multispan4 \hfil  Anion \hfil \\
 &  \multicolumn{1}{c}{$\lambda$} &   \multicolumn{1}{c}{I} &   \multicolumn{2}{c}{I(CH)} &  \multicolumn{1}{c}{$\lambda$} &  \multicolumn{1}{c}{I}&  \multicolumn{2}{c}{I(CH)} & \multicolumn{1}{c}{$\lambda$} &  \multicolumn{1}{c}{I}&   \multicolumn{2}{c}{I(CH)} \\
 & & & nb\tablenotemark{a} & b\tablenotemark{b}& & & nb\tablenotemark{a} & b\tablenotemark{b}& & & nb\tablenotemark{a} & b\tablenotemark{b}\\[5pt]
C$_{96}$H$_{24}$ & 3.255 & 630.6 & 26.3 & & 3.265 & 930.5 & 38.8 & & 3.277 & 1480.5 & 61.7& \\[10pt]

C$_{84}$H$_{24}$ & 3.259 & 215.5 & 18.0 & & 3.272 & 333.6 & 27.8 & & 3.287 & 590.9 & 49.2 &  \\
                 & 3.203 & 303.5 & & 25.3 & 3.213 & 413.0 & & 34.4 & 3.225 & 695.6 & & 58.0 \\[5pt]

C$_{90}$H$_{30}$ & 3.255 & 591.1 & 24.6 & & 3.266 & 1023.2 & 42.6 & & 3.278 & 1579.9 & 65.8& \\
                 & 3.170 & 3.5   & &  0.6 & 3.174 & 4.3    &  & 0.7 & 3.176 & 4.9    &  & 0.8 \\[5pt]

C$_{102}$H$_{26}$\,C$_{2v}$ & 3.256 & 403.8 & 22.4 & & 3.269 & 638.4 & 35.5 & & 3.281 & 1068.8 & 59.4& \\
                     & 3.207 & 235.8 & & 29.5 & 3.214 & 343.8 & & 43.0 & 3.224 & 561.3 & & 70.2 \\[5pt]

C$_{102}$H$_{26}$\,C$_{2h}$\,\tablenotemark{c} &  3.250& 543.6& 24.7& & 3.260& 891.9& 40.5& & 3.272& 1563.6& 71.1 &\\
                             &  3.221&  82.0& & 20.5& 3.231& 142.3& & 35.6& 3.236&  222.5& & 55.6\\[5pt]

C$_{110}$H$_{30}$ & 3.253 & 278.1 & 19.9 & & 3.266 & 465.7 & 33.3 & & 3.280 & 753.0 & 34.2& \\
                  & 3.211 & 400.4 & & 25.0 & 3.217 & 517.3 & & 32.3 & 3.225 & 788.9 & & 49.3 \\[5pt]

C$_{120}$H$_{36}$\,\tablenotemark{c} & 3.249 & 278.6 & 23.2 & & 3.256 &  435.8 & 36.3 & & 3.268 & 704.5 & 58.7& \\
                        & 3.217 & 528.2 & & 22.0 & 3.222 &  532.8 & & 22.2 & 3.230 & 683.7 & & 28.5\\
\hline \\[-10pt]
\end{tabular}
\tablenotetext{a}{non-bay hydrogens;}
\tablenotetext{b}{bay hydrogens;}
\tablenotetext{c}{reducing the FWHM to 15 cm$^{-1}$ to resolve the bay and non-bay H's yields.}
\end{center}
\noindent
\end{table*}
\clearpage

The PAH structures considered here are shown in Figure~\ref{structure}\ and are all variants of the PAH \stref, one of the molecules treated in Paper I. As shown in Figure~\ref{structure}, \stref\, is a compact, highly symmetric (D$_{6h}$), closed edged PAH containing no bay regions and having only solo and duo peripheral hydrogens.  Close inspection of Figure~\ref{structure} shows the following relationships between \stref\, and the PAHs considered here.  The smallest molecule, \sta, results by removing the 12 outermost corner carbons from \stref.  While the symmetry of the molecule remains unchanged, this modification opens the edge structure by introducing 6 bay regions and converts this to a PAH with only duo hydrogens.  Two of the six bay regions are indicated by the arrows in Figure~\ref{structure}.  The conversion to \stb\, is achieved by replacing the 6 carbon atoms comprising the central hexagonal ring of \stref\, with hydrogen atoms.  This too maintains symmetry, while introducing internal hydrogen atoms within the aromatic network.  The structure for the PAH, \stc, is produced by adding 3 carbon atoms to each of two edge rings along one of the symmetry axes.  This adds two rings, reduces the symmetry to C$_{2v}$, introduces 4 bay regions and 6 tertiary hydrogen atoms.  The \std\, isomer is produced by again adding 3 carbon atoms, but this time off a symmetry axis.  This reduces the symmetry further to C$_{2h}$, introduces 2 bay regions and again adds two new rings and 6 tertiary hydrogen atoms.  The modifications that produce \ste\, and \stf, the largest PAHs considered here, introduce bay regions as well as trio and quartet hydrogens to the structure.  The neutral, cation and anion forms for all of these PAHs have been computed and are discussed below.

\clearpage
\begin{figure*}[t!]
   \begin{minipage}[c]{0.33\textwidth}
      \centering \includegraphics[width=\textwidth]{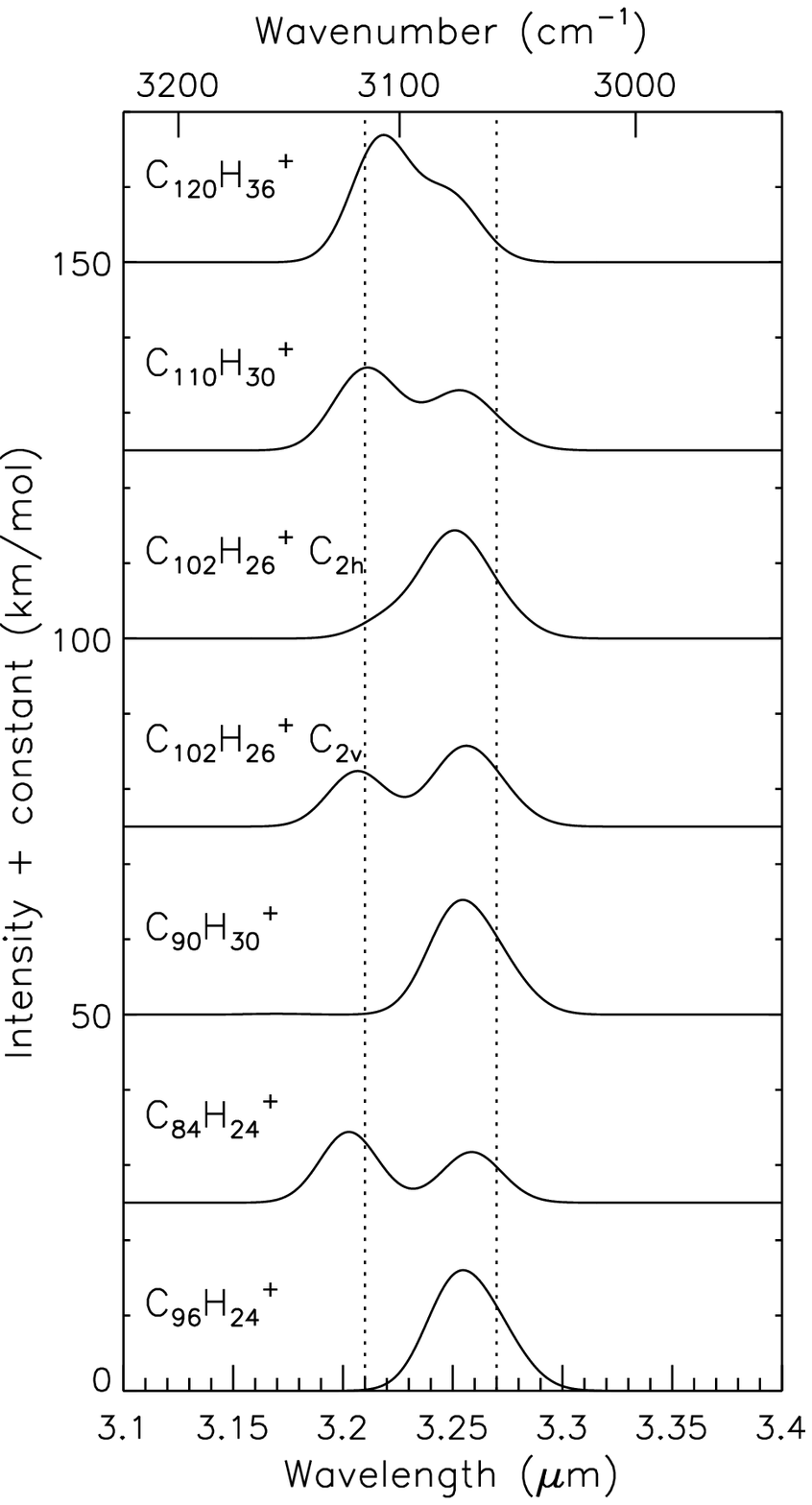}
   \end{minipage}
   \begin{minipage}[c]{0.33\textwidth}
      \centering \includegraphics[width=\textwidth]{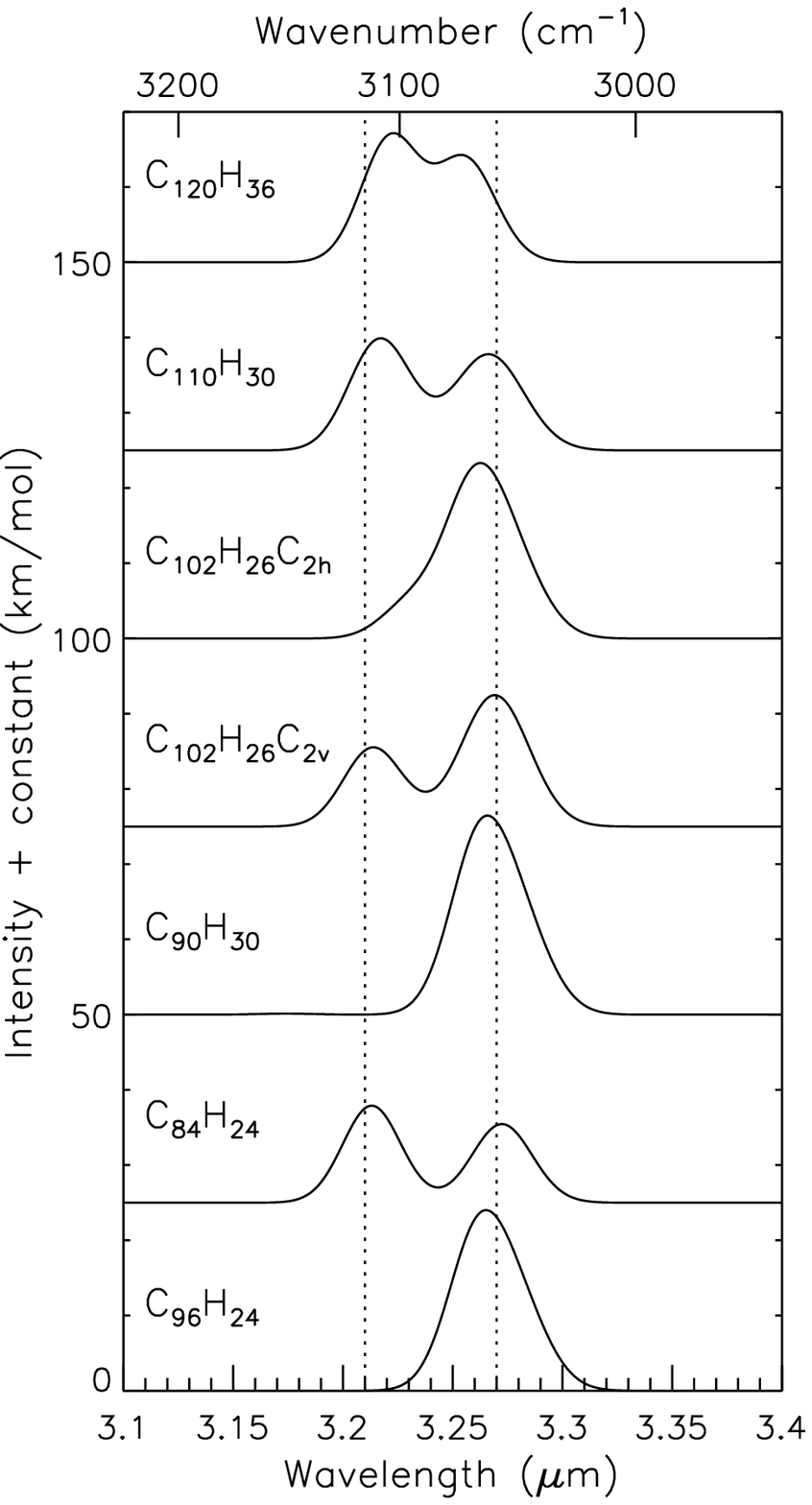}
   \end{minipage}
   \begin{minipage}[c]{0.33\textwidth}
      \centering \includegraphics[width=\textwidth]{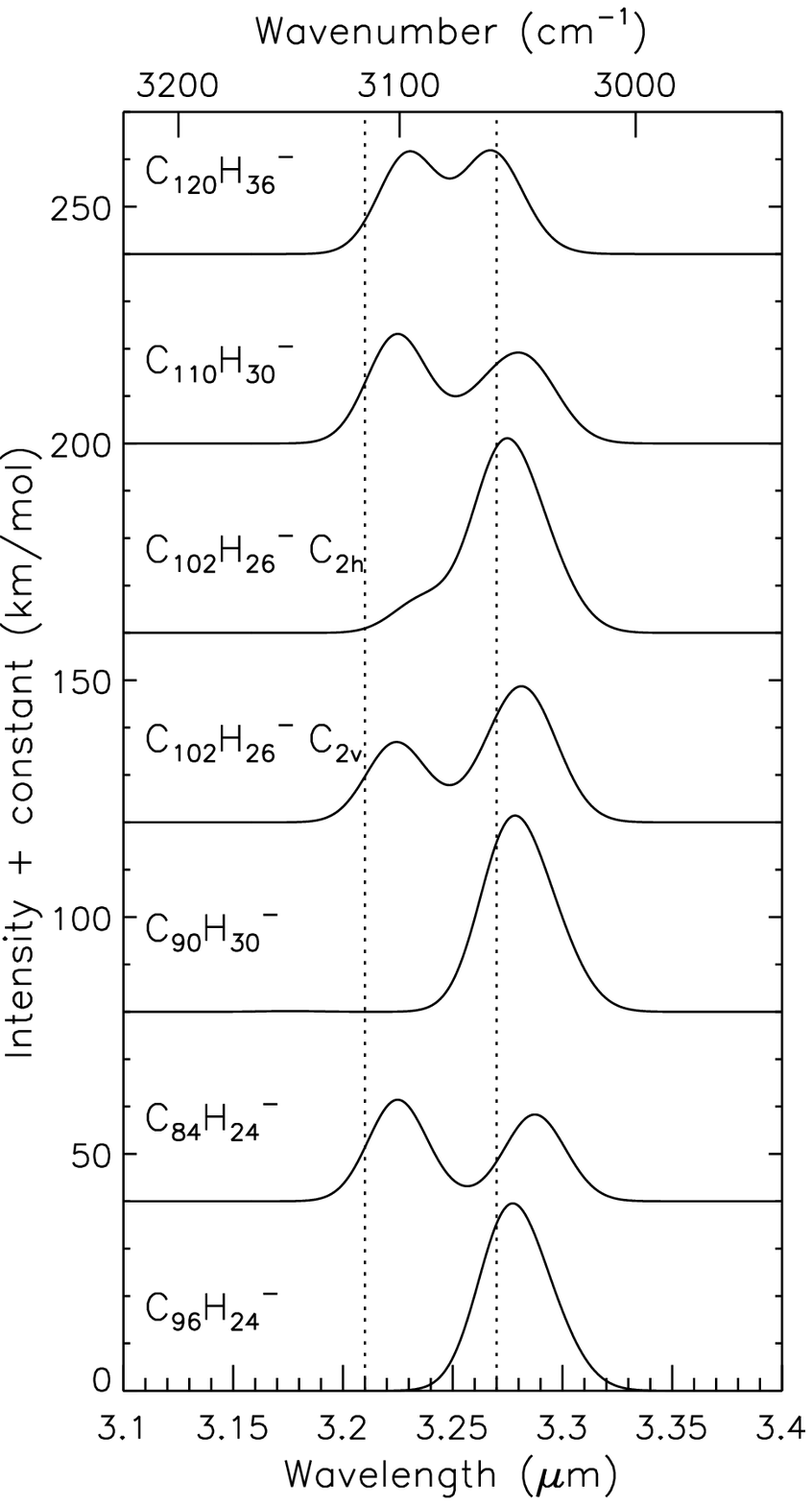}
   \end{minipage}
\caption{The synthetic absorption spectra in the 3 $\mu$m\, region for
the cation, neutral and anion forms of the PAHs shown in
Fig. \ref{structure}. To guide the eye, dotted lines at 3.21 and 3.27
$\mu$m\, are also shown. Note that in contrast with Table \ref{t1}, a FWHM of 30 cm$^{-1}$ is used for all molecules.}
\label{fig_33}
\end{figure*}
\clearpage

\subsection{The CH stretching vibrations (2.5 - 3.5 \mum)}
\label{model_33}

The peak positions of the bands which dominate the CH stretching region for each of the PAHs  are listed in Table~\ref{t1} and the corresponding spectra are shown in Figure~\ref{fig_33}.  

\clearpage
\begin{figure}
  \centering \includegraphics[width=.45\textwidth]{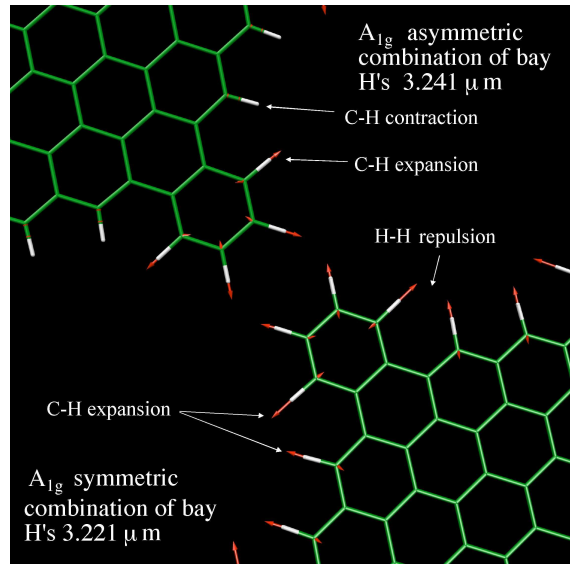}
\caption{ The vibrational motions of the atoms comprising bay and
non-bay edge structures in \stf.  Green represents the carbon
skeleton, white the H atoms and the relative motions are shown with
red arrows.  Note that for a bond contraction, the arrows overlap the
bonds and are a bit hard to see.  The top left structure shows the
atomic motions for the A$_{1g}$ asymmetric vibration.  Here, the
vibrating atoms across the bay are out-of-phase with each other (CH contraction combined with CH expansion).  The
lower structure shows the corresponding motions for the symmetric
combination in which the vibrations are in-phase (both CH bonds stretching at the same time), maximizing the
interaction. }
\label{bayvsnonbay}
\end{figure}

\begin{figure}
  \centering \includegraphics[width=.45\textwidth]{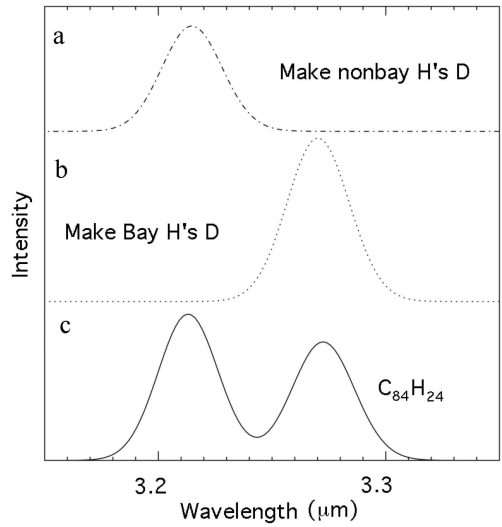}
\caption{Decomposition of the CH$_{str}$ spectrum of \sta.  {\bf a)} The
spectrum when all non-bay hydrogen atoms are replaced by deuterium
atoms. {\bf b)} The spectrum when all bay hydrogen atoms are replaced by
deuterium atoms. {\bf c)} The spectrum of the fully hydrogenated \sta. This
shows that the peak near 3.27 \mum\, is primarily due to non-bay CH
stretching vibrations and the peak near 3.21 \mum\, is largely due to
CH stretching vibrations involving hydrogen atoms across the bay
regions.  See text and Figure \ref{bayvsnonbay} for details.}
\label{sp_bayvsnonbay}
\end{figure}
\clearpage

The spectra in Figure~\ref{fig_33} are striking.  Instead of showing a single strong band peaking near 3.26 \mum\, as is the case for the vast majority of PAHs studied to date, most of the spectra in Figure~\ref{fig_33} have two prominent bands, one peaking near $\sim$3.27 \mum\, and another near $\sim$3.21 \mum.  Comparing the structures shown in Figure~\ref{structure} with the corresponding spectra in Figure~\ref{fig_33} shows that the peak near 3.21 \mum\, is associated with PAHs that have bay region edge structures. Consideration of the specific CH stretching vibrational motions that produce these bands shows that the shorter wavelength (higher frequency) band results from the steric interaction between the hydrogen atoms that are across bay regions.  As illustrated in Figure~\ref{bayvsnonbay}, the stretching vibration of one hydrogen in a bay is hindered by the opposing hydrogen atom, requiring slightly more energy for the transition to occur.  Note that the symmetric combination has a larger shift than the asymmetric combination, but the asymmetric vibration is also shifted with respect to the unhindered CH stretch.  However, all of the asymmetric modes are very weak and therefore essentially make no contribution to the CH stretching region. This interaction is explored further for the molecule, \sta.   The spectrum was computed for two isotopomers of neutral \sta, one form in which all the non-bay hydrogens were replaced by deuterium atoms, the other in which all the bay hydrogens were replaced with deuterium atoms.  The results are shown in Figure~\ref{sp_bayvsnonbay}.  Since the CD stretch is shifted out of this region to $\sim$4.6 \mum, the spectrum is simplified by removing the contributions of CH bands associated with bay or non-bay hydrogens from the region near 3.3 \mum.  Figure~\ref{sp_bayvsnonbay} shows that, when deuterium atoms are substituted for the hydrogen atoms in the bay regions, the band near 3.21 \mum\, in \sta\, disappears while that near 3.27 \mum\, remains.  The opposite occurs when deuterium atoms are substituted for hydrogen at the non-bay positions. This decomposition of the CH stretches clearly shows that the peak near 3.27 \mum\, is due to the non-bay hydrogens and the peak near 3.21 \mum\, is due to the hydrogens across the bay regions.  It is interesting to note that the shift due to the bay structure is, in general, larger than the shift associated with PAH charge, at least for PAHs of this size.

The positions of the bands produced by the CH stretching vibrations of the non-bay hydrogens for the irregular PAHs treated here are similar to those of the large compact PAHs discussed in Paper I, all of which have only non-bay hydrogen atoms.  Table 1 shows that the non-bay CH band in the cations falls between 3.249 and 3.259 \mum, compared with the range of 3.252 to 3.257 \mum\, for the comparably sized PAHs (C$_{78}$H$_{22}$ -- C$_{130}$H$_{28}$) discussed in Paper I.  The CH stretch for the neutral forms of the PAHs shown in Figure~\ref{structure} also span a slightly wider range than those in Paper I and are centered at about the same wavelength.  In this case, the CH stretching band in the irregular PAHs peaks between 3.256 and 3.272 \mum\, compared to 3.264 and  3.267 \mum\, for the earlier sample.   Similarly, the region in which the non-bay CH stretches for the anion forms of these PAHs falls, 3.268 to 3.287 \mum\, also overlaps that of the earlier sample (3.276 to 3.278 \mum). 

The overall band positions and band strength per CH associated with bay and non-bay hydrogens are given in Table~\ref{t1}.    The intensities of the bands produced by the non-bay CH stretching vibrations, are similar to those for the PAHs considered in Paper I.  However, the tendency of the individual CH band strengths of the more symmetric PAHs to grow slightly with PAH size is not reflected in this sample.  Overall molecular structure is clearly important.  The mean value of I(CH) for the 6 PAH cations  considered here is  22.1 km/mol compared with a value that steadily increases from 20.6 to 30.1 km/mol for the comparably sized PAHs in  Paper I. Similarly, the average I(CH) for the neutral PAHs considered here is 36.0 km/mol compared to a range of 36.4 to 43.6 km/mol for the PAHs in Paper I.  Likewise, the average for the anions of the irregular PAHs is 56.4 km/mol versus a steady increase from 61.8 to 65.9 km/mol with size for the compact PAHs.  

Regardless of the somewhat different behavior between I(CH) for the very large PAHs considered in Paper I and the bands produced by the non-bay CH groups in the very large PAHs treated here, these new data reinforce the important differences between the spectroscopy of small PAHs and PAHs which are comparable in size to those which dominate the astronomical mix.  Most significantly, I(CH) for the neutral forms is of the same order as in the ionized forms (within a factor of 1.5). Thus the dramatic reduction in the A value per CH for the CH$_{str}$ upon ionization that is observed for small PAHs is reduced considerably for all of the PAHs considered here and in Paper I.  The trend of intensity increase as PAH charge form changes from cation to neutral to anion holds for all the PAHs considered here and in Paper I.

\clearpage
\begin{deluxetable}{l@{\hspace{5pt}}r@{\hspace{8pt}}r@{\hspace{15pt}}r@{\hspace{8pt}}r@{\hspace{15pt}}r@{\hspace{8pt}}r}
\tablecaption{\label{t2} The 6-9 $\mu$m band position maxima ($\lambda$, in
$\mu$m) and total intensity (I).  The intensities are in km/mol. }
\tablehead{\colhead{Molecule} & \multicolumn{2}{l}{Cation} & \multicolumn{2}{l}{Neutral} & \multicolumn{2}{l}{Anion} \\
 &  \multicolumn{1}{c}{$\lambda$} &  \multicolumn{1}{c}{I} &  \multicolumn{1}{c}{$\lambda$} &  \multicolumn{1}{c}{I}&   \multicolumn{1}{c}{$\lambda$} &  \multicolumn{1}{c}{I}}
\startdata
 C$_{96}$H$_{24}$
 &    8.938&      62.9\\
 &    8.401&    1023.3& 8.592&      77.4& 8.463&    1352.4\\
 &     & & 8.286&       4.7\\
 &    7.782&     469.8& 7.846&      56.6& 7.833&     932.0\\
 &    7.563&    1816.1& 7.590&      16.9& 7.675&    1451.7\\
 &    & & 7.404&      13.4\\
 &    & & 7.054&      20.5& 7.151&     194.5\\
 &    6.827&     997.8& 6.823&      43.8& 6.841&     313.1\\
 &     & & 6.471&      41.6& 6.497&     292.7\\
 &    6.345&    2718.2& 6.265&     105.7& 6.354&    1253.4\\[5pt]

C$_{84}$H$_{24}$ 
&     8.932&      11.7& 8.941&       1.1\\
&      & & 8.535&       6.2\\
&     8.034&    1192.5& 7.928&     170.4& 8.065&     866.6\\
&     7.588&    1076.6& 7.468&      46.2& 7.646&    2376.9\\
&     7.153&      57.5& & & 7.244&      54.0\\
&     6.855&      19.0& 6.878&      37.2& 6.882&      53.9\\
&     6.624&      47.0& 6.727&       8.9& 6.693&     126.8\\
&     6.351&    1879.0& 6.244&      82.0& 6.346&    1270.5\\[5pt]

C$_{90}$H$_{30}$  
 &    8.695&     475.7& 8.720&      45.7& 8.746&     446.1\\
 &    8.415&     474.4& 8.364&      36.2& 8.451&     429.4\\
 &    7.933&     628.1& 8.055&      91.7& 8.037&     735.2\\
 &    7.590&     377.9& 7.762&       3.3& 7.612&     401.9\\
 &    7.330&     312.4& 7.412&      68.1& 7.369&     314.5\\
 && & & & 7.044&      81.6\\
 &    6.821&     673.0& 6.821&      36.7& 6.817&     242.8\\
 &    6.411&    1409.1& 6.405&     103.0& 6.445&     790.6\\
 &     & & 6.267&      83.8\\[5pt]

C$_{102}$H$_{26}$C2v 
&     8.941&      12.4& 8.951&      32.9& 9.157&      13.7\\
&     8.431&     755.5& 8.372&      23.2& 8.573&     601.8\\
&     8.000&     665.1& 8.179&      41.3& 8.300&     194.3\\
&     7.785&    1580.1& 7.634&     123.9& 7.868&    2270.4\\
&     7.023&     333.1& 7.141&      59.2& 7.361&     172.6\\
&     & & & & 7.103&     148.0\\
&     6.788&     249.6& 6.853&      59.8& 6.889&      81.5\\
&      & & & &6.599&     225.9\\
&     6.416&    1978.6& 6.329&     191.8& 6.386&     881.7\\[5pt]

 C$_{102}$H$_{26}$C2h
&     8.947&      51.9& 8.962&      40.2\\
&     8.480&    1082.0& 8.449&      58.6& 8.592&    1700.0\\
&     7.828&    1740.7& 7.928&      31.2& 7.838&    1837.9\\
&    & & 7.541&     126.2\\
&     7.375&     213.8& 7.308&      51.2& 7.379&     424.4\\
&     6.999&     109.2& 6.964&      62.9\\
&     6.813&     301.3& 6.775&      37.1& 6.841&    1218.7\\
&     6.508&     492.1& 6.693&      46.0& 6.544&     844.7\\
&     6.310&     342.2& 6.292&     241.2& 6.356&     753.4\\[5pt]

C$_{110}$H$_{30}$
&      & & 8.699&      24.5& 8.774&     198.7\\
&     8.258&     804.9& 8.275&      72.2& 8.402&     118.3\\
&     7.988&     769.7\\
&     7.819&    1784.5& 7.759&     104.9& 7.901&    2941.7\\
&     & & & & 7.389&     241.4\\
&     7.049&     403.3& 7.055&     112.0& 7.146&      94.7\\
&     & & & & 6.853&      83.0\\
&     6.758&     196.5& 6.768&     168.9& 6.715&     162.6\\
&    & & & & 6.567&     157.3\\
&     6.407&    2197.3& 6.321&     186.2& 6.383&     914.8\\[5pt]

C$_{120}$H$_{36}$
&     8.907&       3.4& 8.911&       8.3& 8.937&     102.9\\
&     8.447&     266.9& 8.368&      23.3\\
&     8.119&    1150.3& 8.145&      48.7& 8.183&     833.7\\
&     7.830&     904.3& 7.838&     113.2& 7.870&    1642.6\\
&     7.552&    1710.9& 7.449&      38.1& 7.603&    2295.3\\
&     7.189&     530.3& 7.153&      23.8& 7.094&     225.6\\
&      & & 6.973&      49.5\\
&     6.734&     731.0& 6.724&     340.6& 6.776&     686.9\\
&     6.411&    2722.5& 6.292&     131.3& 6.386&    1859.1\\
\enddata
\end{deluxetable}
\clearpage

\begin{figure*}[t!]
   \begin{minipage}[c]{0.33\textwidth}
      \centering \includegraphics[width=\textwidth]{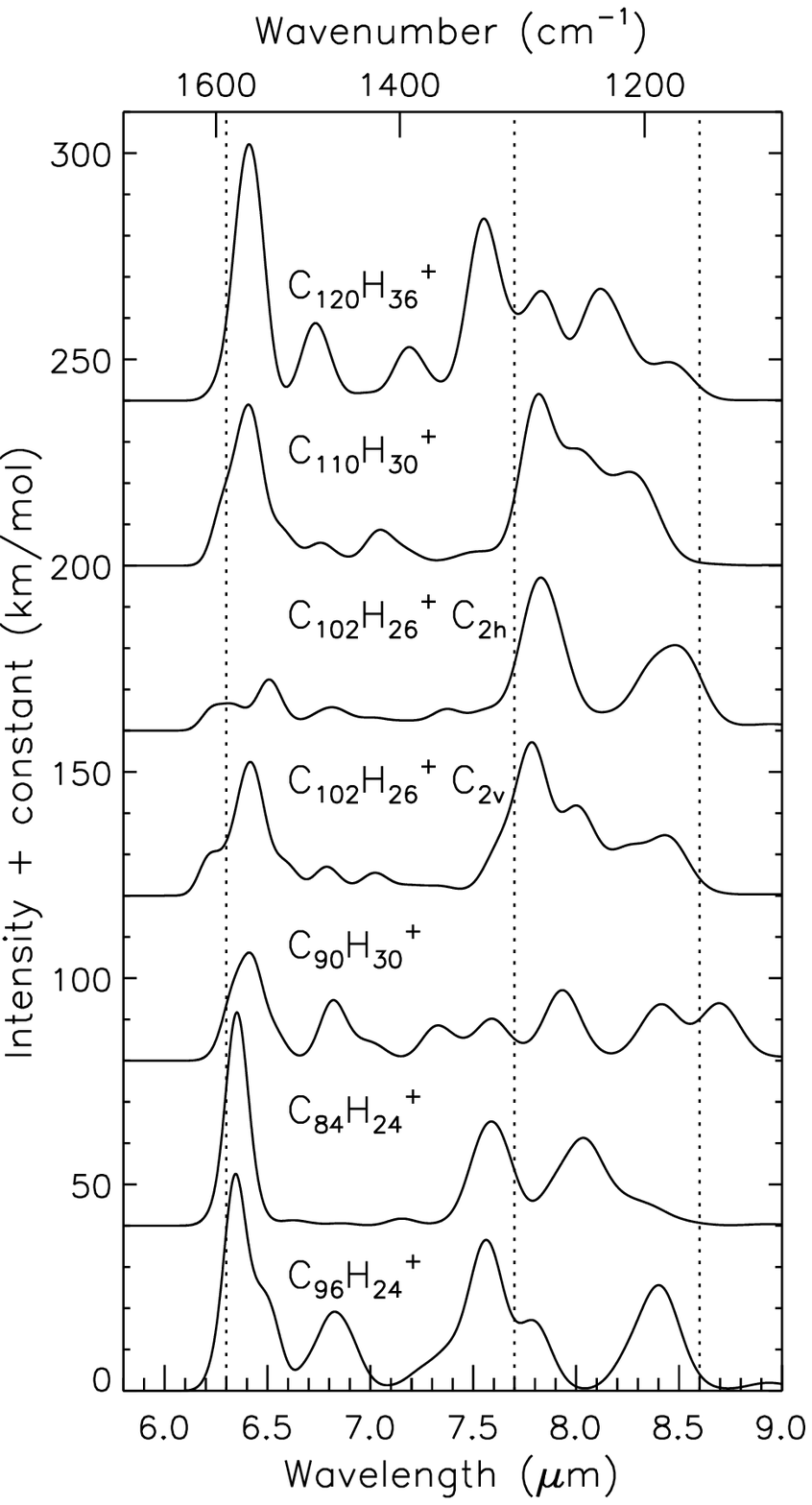}
   \end{minipage}
   \begin{minipage}[c]{0.33\textwidth}
      \centering \includegraphics[width=\textwidth]{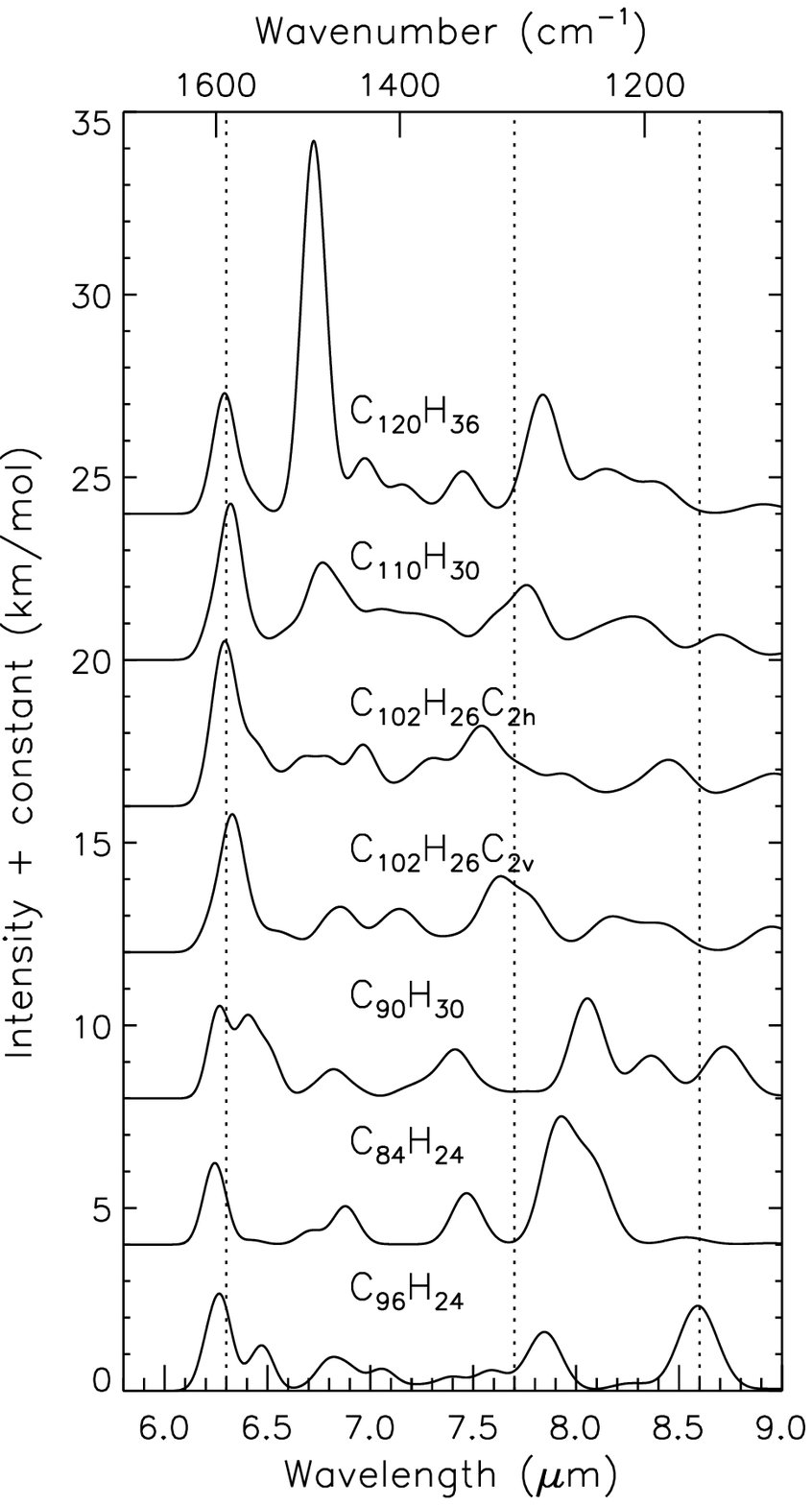}
   \end{minipage}
   \begin{minipage}[c]{0.33\textwidth}
      \centering \includegraphics[width=\textwidth]{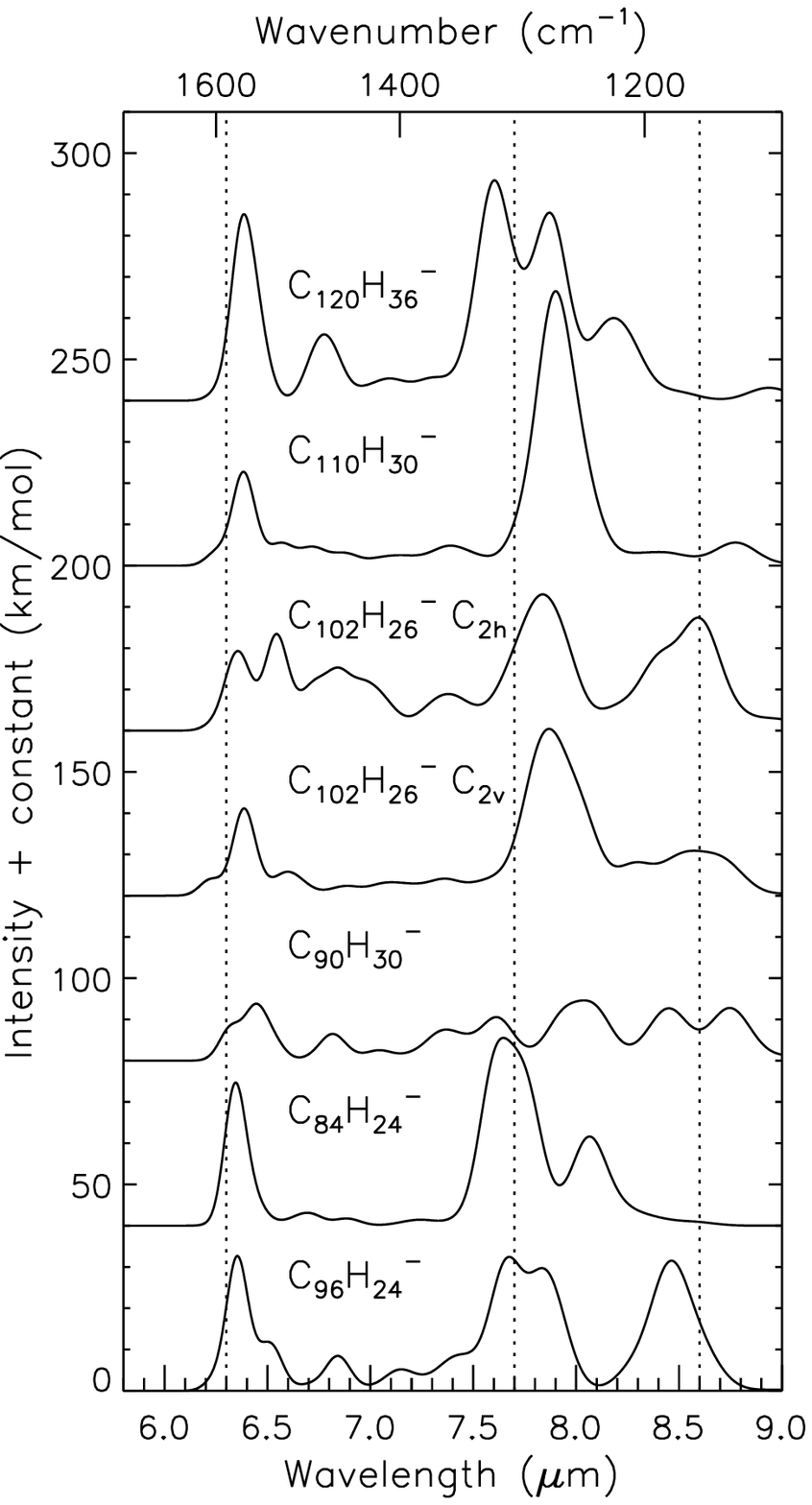}
   \end{minipage}
\caption{The synthetic absorption spectra in the 6 to 9 $\mu$m\,
region for the cation, neutral and anion forms of the PAHs shown in
Figure \ref{structure}. To guide the eye, dotted lines at 6.3, 7.7 and
8.6 $\mu$m\, are also shown. }
\label{fig_62}
\end{figure*}
\clearpage

\subsection{The CC stretching and CH in-plane bending vibrations (5-9 \mum)}
\label{model_62}

The 5-9 \mum\, region of the spectra for the neutral, cation, and anion forms of the PAHs considered here are shown in Figure~\ref{fig_62} with corresponding band positions and integrated band strengths listed in Table~\ref{t2}.  The bands in this wavelength range correspond to CC stretching and CH in-plane bending vibrations.  In considering the spectra in Figure~\ref{fig_62}, one is struck by the prominent band at 6.725 \mum\, for neutral \stf.  Analysis of the vibrational motions shows that this band is produced primarily by the CH in-plane bending motions of the 24 quartet hydrogen atoms on the protruding rings in \stf.  The band is visible in the spectrum of \ste, the only other PAH in this set which contains protruding rings, but it is much weaker since this PAH has only 8 quartet hydrogen atoms, a number comparable to the number of solo, duo, and trio hydrogens.  This quartet hydrogen mode is much less obvious in the ions.  Note that the integrated band strengths for the neutral PAHs is at least 10 times smaller than those of the corresponding cations and anions.  The significant intensity enhancement of the bands in the 6-9 \mum\, region upon PAH ionization is consistent with that found for all PAHs studied to date \citep[][and ref. therein]{Szczepanski:labcations;93, Langhoff:neutionanion:96, Kim:gasphasepyrenecation:01, Hudgins:tracesionezedpahs:99}.  Since the bands between 6 and 9 \mum\, in the spectra of neutral PAHs are an order of magnitude weaker than those in the corresponding cations and anions, the discussion in this section focuses only on the cation and anion spectra.  We should note that for the nitrogenated PAHs (PANHs) the intensity of the emission from the neutrals is not insignificant when compared with the cations or anions, and hence for PANHs the neutrals would need to be included in the discussion, which is not the case for the pure PAHs.

The pure CC stretching region ($\sim$6.2 to 6.4 \mum) in the spectra of the PAH cations and anions shown in Fig.~\ref{fig_62} are rather similar and, perhaps surprisingly, show much less variation than do the spectra of the more symmetric, regular PAHs described in Paper I.  With the exception of the C$_{2h}$ forms of C$_{102}$H$_{26}$\,$^+$ and C$_{102}$H$_{26}$\,$^-$, the spectra for all of these PAHs have a prominent CC stretching band peaking between 6.346 and 6.445 \mum.  

The features that fall between 7 to 9 \mum\, in the spectra shown in Fig.~\ref{fig_62} are produced by CC stretching and CH in-plane bending vibrations.  Vibrations involving CC stretching motions coupled with CH in-plane bending modes generally produce bands in the 7 to 8 \mum\, region while CH in-plane bending vibrations usually produce features between 8 and 9 \mum.  Overall, the 7 to 9 \mum\, spectra of the cations and anions resemble one another, but not as closely as the spectra of the similarly sized PAHs treated in Paper I. There we pointed out that a study of the modes suggested that the shape of the edge might contribute to the differences in this region of the spectrum as they do in the CH$_{str}$ and CH$_{oop}$ regions. For example, there seems to be a jump from 7.6 to 7.8 \mum\, at C$_{102}$ for both the cations and anions.  A similar shift is found for the PAHs at C$_{110}$ in Paper I. However, as Fig.~\ref{fig_62} shows, the spectrum of \stf, a highly symmetric, yet significantly less compact PAH than all the others considered, seems to present an exception. Here the strongest peak falls near 7.6 \mum, close to the position of the strong band in the spectra of the smaller PAHs  \sta\, and \stref.  Apparently the ``extra" protruding rings in \stf\, do not couple strongly enough with the C$_{96}$ core to shift this band. Finally, it is interesting to note that none of the spectra shown in Figure~\ref{fig_62} have a strong 8.6 \mum\, band except for that of the C$_{2h}$ form of C$_{102}$H$_{26}$.  This is in sharp contrast with the spectra of the large PAHs considered in Paper I.  The 8.6 \mum\, band appears to grow in intensity with the increasing size of large symmetric, compact PAH species, with some of those larger than \stref\, having rather strong, prominent bands near 8.6 \mum. However, this is not observed for the  irregular PAHs considered here.

\clearpage
\begin{figure*}[t!]
   \begin{minipage}[c]{0.33\textwidth}
      \centering \includegraphics[width=\textwidth]{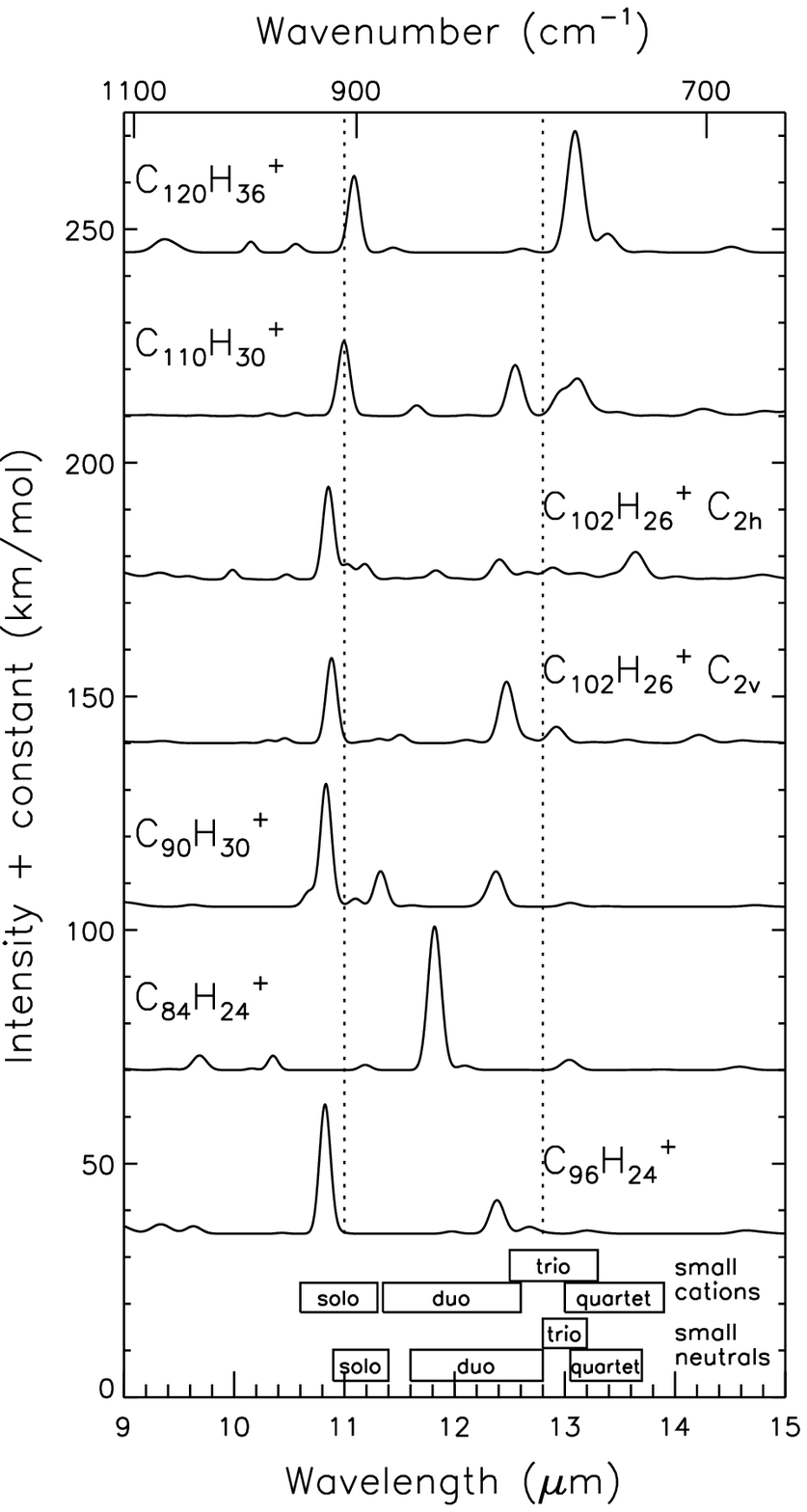}
   \end{minipage}
   \begin{minipage}[c]{0.33\textwidth}
      \centering \includegraphics[width=\textwidth]{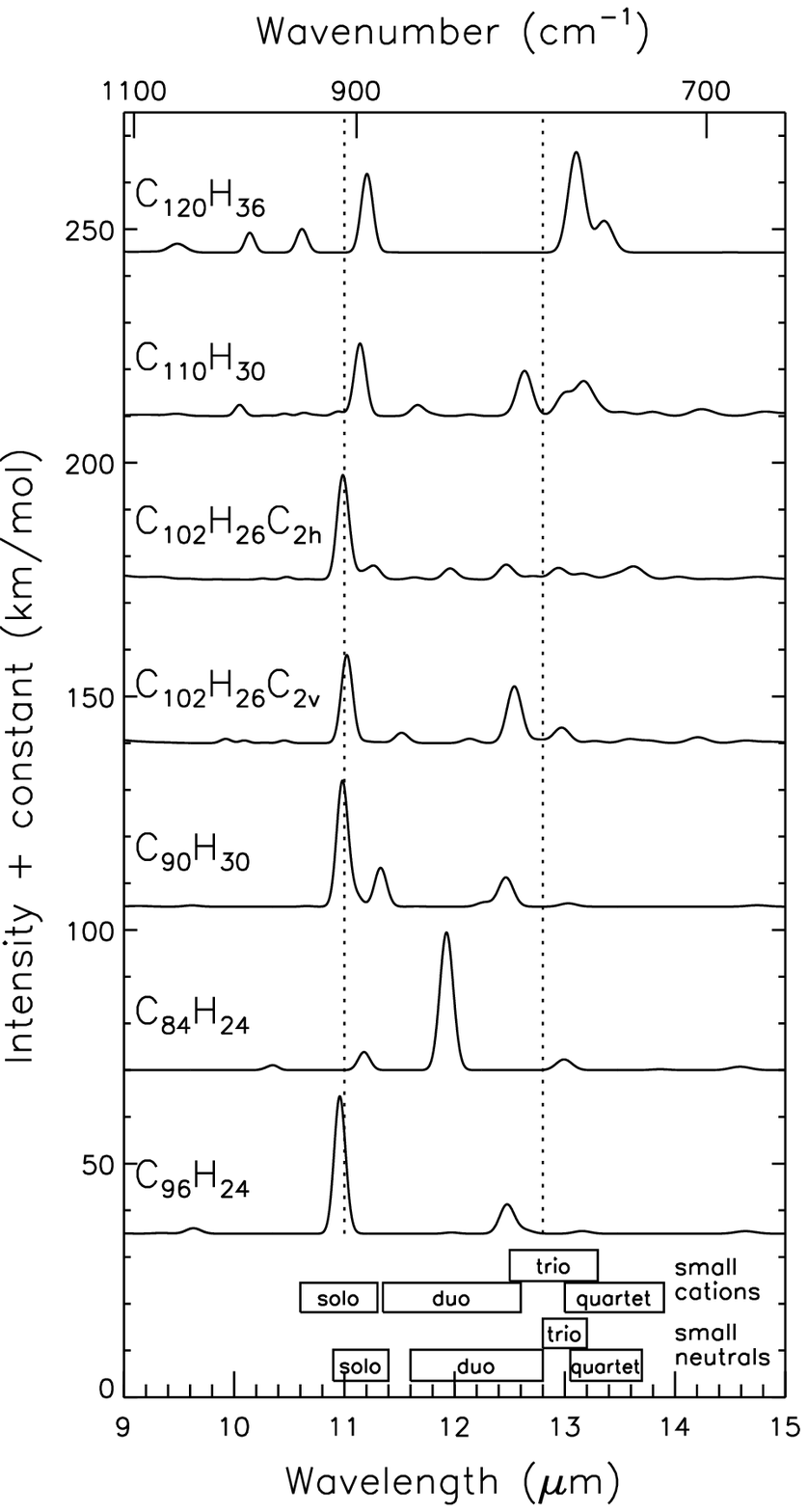}
   \end{minipage}
   \begin{minipage}[c]{0.33\textwidth}
      \centering \includegraphics[width=\textwidth]{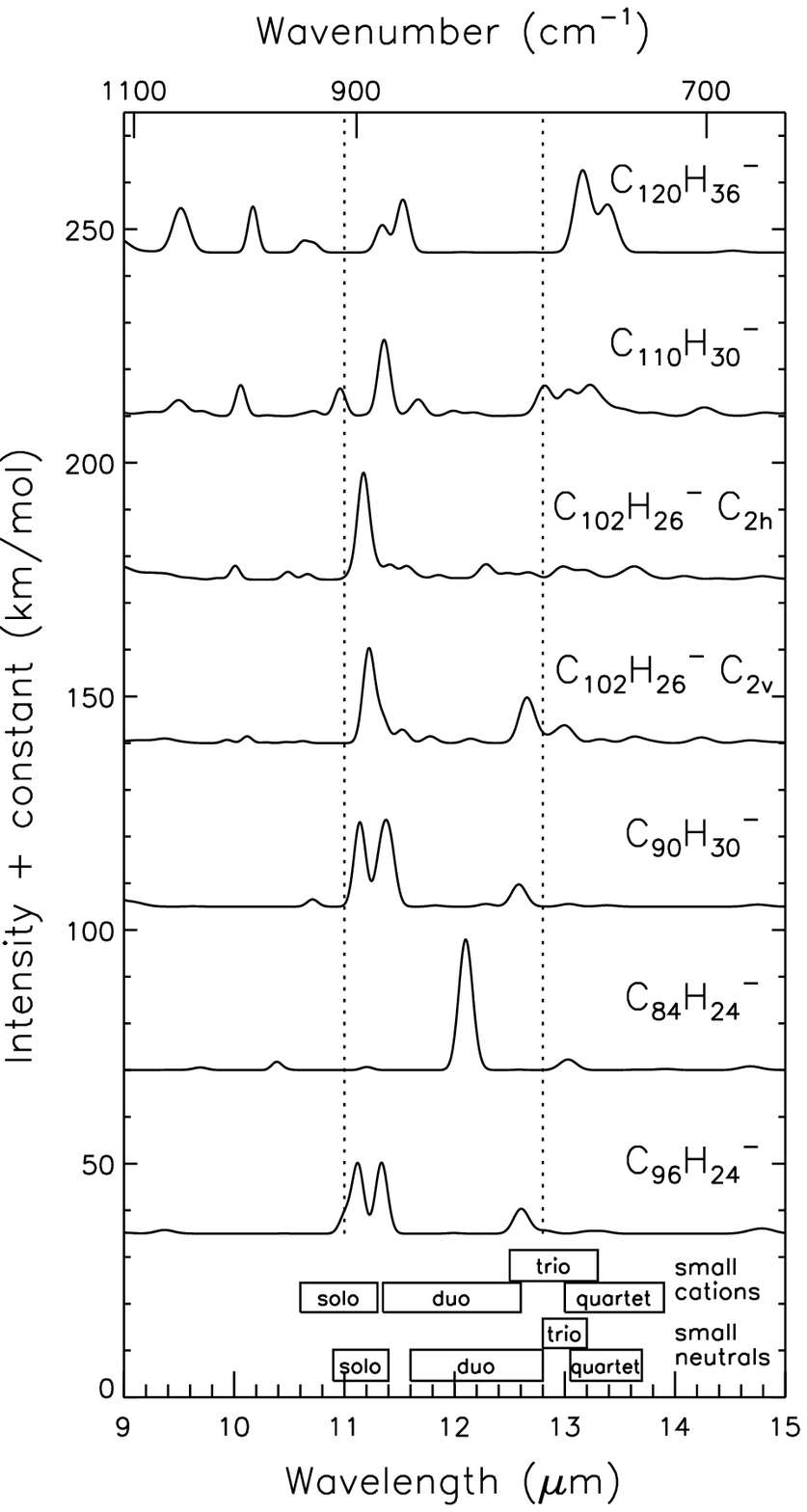}
   \end{minipage}
\caption{The synthetic absorption spectra in the 9 to 15 $\mu$m\,
region for the cation, neutral and anion forms of the PAHs shown in
Figure \ref{structure}. To guide the eye, dotted lines at 11.0 and
12.9 $\mu$m\, are also shown. The boxes in the lower frame indicate
the wavelength regions associated with the CH$_{oop}$
vibrations for different types of adjacent hydrogen atoms determined
from matrix isolated spectroscopy of neutral and cationic small PAHs
\citep{Hudgins:tracesionezedpahs:99, Hony:oops:01}. }
\label{fig_112}
\end{figure*}
\clearpage

\begin{deluxetable}{l@{\hspace{5pt}}r@{\hspace{8pt}}r@{\hspace{12pt}}r@{\hspace{8pt}}r@{\hspace{12pt}}r@{\hspace{8pt}}r}
\tablecaption{\label{t3} The 9-15 $\mu$m band position maxima
($\lambda$, in $\mu$m) and total intensity (I, in km/mol). The
corresponding number of solo, duo, trio and quartet hydrogens are
given in parenthesis under each formula. For PAHs with {\it any} band
stronger than 10km/mol, all charge states are listed. If the bands for
the cation, neutral and anion state of a PAH all fall below 10km/mol
they are not included. }
\tablehead{Molecule & \multicolumn{2}{l}{Cation} &\multicolumn{2}{l}{Neutral} & \multicolumn{2}{l}{Anion} \\
 &  \multicolumn{1}{c}{$\lambda$} &  \multicolumn{1}{c}{I} &  \multicolumn{1}{c}{$\lambda$} &  \multicolumn{1}{c}{I}&   \multicolumn{1}{c}{$\lambda$} &  \multicolumn{1}{c}{I}}
\startdata
C$_{96}$H$_{24}$& 14.648&       9.5& 14.639&       5.7& 14.786&      13.9\\
\,\,\,(12:12:0:0) & 13.203&       8.6& 13.158&       6.0& 13.293&      10.3\\
 &   12.677&      17.4& & & 12.606&      65.8\\
 &   12.385&      76.9& 12.477&      74.2\\
 &   11.975&       5.3& 11.972&       2.7& 11.993&       1.6\\
 &    & & & & 11.338&     162.7\\
 &   10.825&     299.0& 10.959&     316.4& 11.120&     199.6\\
 &   10.433&       1.9\\
 &    9.630&      29.7& 9.630&      21.4\\
 &    9.334&      49.3& 9.341&       3.2& 9.369&      18.9\\[5pt]

C$_{84}$H$_{24}$ &    14.586&       8.1& 14.590&       7.9& 14.684&       9.3\\
\,\,\,(0:24:0:0) &    13.883&       1.6& 13.864&       2.0& 13.914&       3.0\\
&    13.041&      23.5& 12.994&      24.4& 13.029&      25.5\\
&    12.092&      10.0& 11.927&     314.0& 12.101&     301.1\\
&    11.819&     331.6\\
&    11.191&      12.0& 11.178&      41.3& 11.204&       7.1\\
&    10.353&      33.0& 10.350&      13.4& 10.387&      20.6\\
&     9.686&      53.7& & & 9.691&      10.0\\[5pt]

C$_{90}$H$_{30}$ &   14.732&       3.3& 14.745&       3.1& 14.749&       4.6\\
\,\,\,(18:12:0:0) &   13.050&       8.6& 13.031&       7.6& 13.036&       6.1\\
 &   12.376&      88.2& 12.466&      76.3& 12.583&      50.8\\
 &   11.329&      81.3& 11.329&      89.1& 11.378&     248.6\\
 &   11.100&      17.5\\
 &   10.833&     321.7& 10.983&     311.2& 11.141&     193.1\\
 &    & & 10.659&       1.4& 10.711&      16.7\\
 &    9.619&       7.3& 9.616&       4.9& 9.626&       1.4\\[5pt]

 C$_{102}$H$_{26}$C$_{2v}$ & 14.616&       8.6& 14.648&       8.8& 14.686&       8.2\\
\,\,\,(9:12:6:0)  & 14.219&      19.0& 14.207&      13.4& 14.239&      12.9\\
&    13.565&       9.4& 13.592&      14.7& 13.637&      18.3\\
&    13.261&       2.3& 13.270&       4.6& 13.324&       8.7\\
&    12.923&      37.9& 12.972&      37.9& 12.995&      48.7\\
&    12.472&     149.6& 12.542&     134.8& 12.657&     113.8\\
&    12.109&       8.3& 12.136&      10.0& 12.146&      10.4\\
&     & & & &11.780&      15.3\\
&    11.507&      19.9& 11.519&      25.1& 11.525&      29.6\\
&    11.316&      13.2\\
&    10.884&     193.7& 11.023&     206.0& 11.225&     269.1\\
&    10.460&      11.6& 10.455&       6.1& 10.476&       2.4\\
&    10.310&       6.7& & & 10.288&       1.4\\
&    10.092&       1.1& 10.091&       6.1& 10.117&      15.8\\
&      & & 9.924&      10.9& 9.936&       7.9\\
&     9.348&      12.7& & & 9.367&      29.8\\[5pt]

C$_{102}$H$_{26}$C$_{2h}$ & 14.793&      15.1& 14.751&       8.7& 14.791&       9.2\\
\,\,\,(12:8:6:0) & 14.011&       7.5& 14.029&       6.6& 14.083&       8.5\\
&    13.643&      75.5& 13.622&      42.3& 13.630&      41.3\\
&    13.132&      17.3& 13.158&      13.4& 13.168&      22.3\\
&    12.892&      31.9& 12.943&      28.9& 12.985&      34.6\\
&    12.663&      16.6& 12.711&       6.3& 12.665&      18.0\\
&    12.407&      49.5& 12.469&      36.4& 12.483&      12.7\\
&    & & & &12.287&      37.8\\
&    11.832&      27.8& 11.959&      26.9& 11.855&      11.2\\
&    11.472&       3.8& 11.636&       5.6& 11.563&      34.6\\
&     & & & & 11.412&      31.3\\
&    11.184&      36.7& 11.264&      39.7& 11.173&     282.3\\
&    11.025&      28.1& 10.987&     255.8\\
&    10.855&     221.9& 10.661&       1.0& 10.666&      12.5\\
&    10.476&      11.8& 10.477&       6.2& 10.488&      18.0\\
&     9.985&      24.8& 10.259&       2.6& 10.009&      32.6\\
&     9.575&      13.7& 9.564&       3.0& 9.863&       3.3\\
&     9.325&      41.8& 9.301&      19.7\\[5pt]

 C$_{110}$H$_{30}$ &  14.815&      13.2& 14.819&      12.2& 14.824&       8.2\\
\,\,\,(8:8:6:8)&    14.255&      21.2& 14.241&      18.5& 14.265&      21.7\\
&    13.820&       1.9& 13.795&      10.0& 13.787&       7.7\\
&    13.472&       8.6& 13.508&       8.3& 13.229&     109.2\\
&    13.113&     144.0& 13.172&     141.9& 13.036&      54.6\\
&    12.550&     116.1& 12.634&     105.2& 12.819&      71.5\\
&    12.124&       2.8& 12.136&       3.8& 12.167&       8.6\\
& & & & &11.989&      12.4\\
&    11.656&      24.6& 11.666&      29.7& 11.670&      40.1\\
&    10.996&     184.0& 11.142&     167.4& 11.360&     176.2\\
&    10.746&       1.6& 10.946&       8.5& 10.961&      63.2\\
&    10.564&       7.3& 10.634&       9.7& 10.719&      16.1\\
&    10.316&       6.3& 10.456&       5.5& 10.303&       1.8\\
&    10.053&       1.2& 10.050&      25.7& 10.059&      74.3\\
&     9.685&       3.1& 9.474&      12.1& 9.700&      14.8\\
&     & & & & 9.497&      78.0\\
&     9.229&      13.7& 9.192&      11.0& 9.257&      21.2\\[5pt]

C$_{120}$H$_{36}$ &   14.510&      13.1& & & 14.524&       4.4\\
\,\,\,(12:0:0:24)  &   13.387&      41.2& 13.356&      67.0& 13.387&     105.4\\
&    13.092&     279.8& 13.104&     233.1& 13.163&     192.9\\
&    12.617&       8.9& & & 12.077&       1.1\cr
&    11.442&      12.3& & & 11.530&     124.3\cr
&    11.088&     178.4& 11.206&     179.9& 11.344&      60.9\cr
&    10.560&      21.6& 10.616&      54.9& 10.641&      44.3\cr
&    10.151&      24.6& 10.142&      44.9& 10.170&     104.5\cr
&     9.369&      85.2& 9.484&      45.9& 9.516&     199.2
\enddata
\end{deluxetable}
\clearpage

\subsection{The CH out-of-plane bending vibrations (9-15 \mum)}
\label{model_112}

The 9 to 15 \mum\, region of the spectra for the neutral, cation, and anion forms of the irregular PAHs are shown in Fig.~\ref{fig_112} and the peak wavelengths and integrated band strengths of the most significant bands in these spectra are summarized in Table~\ref{t3}.  The bands in Fig.~\ref{fig_112} correspond to CH out-of-plane bending vibrations (CH$_{oop}$).  The lower frames of the figure indicate the regions typically associated with the CH$_{oop}$ bands produced by solo, duo, trio, and quartet hydrogens \citep[C$_n \le$ 32)][]{Hony:oops:01}.  In contrast with the bands in the 5 to 9 \mum\, CC stretching and CH in-plane bending region, the intensities of the CH$_{oop}$ bands for the cations, anions, and neutral PAHs are similar.  While this parallels the intensity behavior in the spectra of the compact, symmetric PAHs discussed in Paper I, the similarities stop there.  Since these irregular PAHs contain varying amounts of solo, duo, trio and quartet hydrogens while the compact PAHs contain only solo and duo hydrogens, the spectra are quite different and variable.

For these PAHs, the CH$_{oop}$ band for the solo hydrogens falls between 10.983 and 
11.206 \mum\, for the neutral forms, between 10.833 and 11.088 \mum\, for the cations, and between 11.141 and 11.530 \mum\, for the anions.  Within each charge group, the peak position tends to shift to slightly longer wavelength with increasing PAH size.  As shown in Figure~\ref{fig_112}, these all fall within the range expected for solo edge hydrogens.

Because of the edge variations in this sample, only one molecule, \stb, has solo and duo hydrogens exclusively.  Its spectrum is similar to that of \stref\, and the compact PAHs considered in Paper I, molecules which have only solo and duo hydrogens.  Although both isomers of C$_{102}$H$_{26}$ have solo, duo, and trio hydrogens, their spectra are surprisingly different between 11.5 and 14 \mum, the region associated with duo and trio CH$_{oop}$ vibrations.  While the C$_{2v}$ and C$_{2h}$ forms have 12 and 8 duo hydrogens respectively, the C$_{2v}$ isomer has a prominent band in the duo region in all charge forms while the C$_{2h}$ isomer shows only a weak feature.    \ste, on the other hand, has comparable numbers of solo, duo, trio, and quartet hydrogens.  The double-humped broad bands centered near 13 \mum\, in the spectra of the cation and neutral forms are produced by overlapping duo, trio, and quartet bands.  These apparently fall closer together in the anion, producing a broad, three peaked band centered near 13.2 \mum.  These vibrations in \ste\, are discussed further in the following paragraph.  \stf\, has only solo and quartet hydrogens, giving rise to two prominent bands falling in the regions expected for these CH$_{oop}$ vibrations. Note that the bands at 9.5, 10.2, and 10.6 \mum\, in \stf\,  are all related to CC stretching vibrations and deformations of the benzene rings containing the quartet hydrogens.   One sees a hint of them in \ste\,  which also has quartet               
hydrogens. 

\clearpage
\begin{figure}
  \centering \includegraphics[width=.45\textwidth]{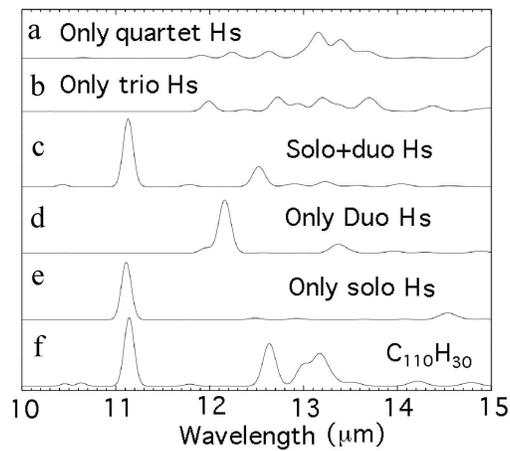}
\caption{Decomposition of the CH$_{oop}$ spectrum of \ste. {\bf a)} The
spectrum when all solo, duo, and trio hydrogens are replaced by
deuterium.  {\bf b)} The spectrum when all solo, duo, and quartet hydrogens
are replaced with deuterium. {\bf c)} The spectrum when all trio and quartet
hydrogens are replaced with deuterium. {\bf d)} The spectrum when all solo,
trio, and quartet hydrogens are replaced by deuterium. {\bf e)} The spectrum
when all duo, trio, and quartet hydrogens are replaced with
deuterium. {\bf f)} The spectrum of the fully hydrogenated species. Note that these spectra are plotted on the same intensity scale and hence can be directly compared with each other.}
\label{sp_oops}
\end{figure}
\clearpage

To gain more insight into these vibrations, the CH$_{oop}$ modes for \ste\, were computationally separated.  \ste\, was chosen because it has comparable numbers of solo (8), duo (8), trio (6), and quartet (8) hydrogens.  The resulting spectra are shown in Figure~\ref{sp_oops}.  The lowest spectrum is that of the parent while the others correspond to cases in which some of the hydrogen atoms have been converted to deuterium atoms.  When all but the solo hydrogen atoms are changed to deuterium atoms (i.e. only solo H atoms remain), the solo peak is very close to its position in the parent, but has lost some intensity. In the case in which only duo hydrogen atoms are present, the strong duo peak falls at $\sim$12.2 \mum\, and is shifted from its position in the parent which is presumably responsible for the 12.6 \mum\, peak in the parent spectrum.  This spectrum looks very much like that of \sta, which has only duo hydrogens.  In the case in which both solo and duo hydrogens are present on the C$_{110}$ structure (`solo+duo H'), the coupling of these hydrogen modes is clearly visible, with the duo band shifting to longer wavelength and losing intensity to the solo hydrogen band.  Converting all but the trio or quartet hydrogen atoms to deuterium atoms (`only trio H' or `only quartet H') shows that the trio and quartet modes exhibit a set of weaker bands which overlap.  There is also some overlap with the duo band if the duo bands are shifted to longer wavelength due to solo-duo coupling. 

This analysis shows that the duo band falls near 12.2 \mum\, when duo hydrogen vibrations are uncoupled from solo bands. While the quartet bands are not significantly coupled to the other bands, there is some coupling between the duo and trio bands.  This coupling is responsible       
for the small shift in the duo band observed in solo+duo H's (Fig.~\ref{sp_oops}c) compared        
with that in all H case (Fig.~\ref{sp_oops}f).  As this coupling is smaller than the solo-duo      
coupling we do not consider it in detail.  We should note that in \ste\, the trio H's are  adjacent to the duo hydrogens, while in other structures they can be adjacent to solo hydrogens, so the coupling between duo and trio hydrogens can be system dependent.      
Nevertheless, this might offer the possibility of identifying PAHs containing duo hydrogens and the charge on the PAH. However, as discussed above, duo bands are shifted to, and overlap with, the red edge of the trio and quartet bands due to coupling with solo hydrogens. This work suggests that bands at 13 \mum\, and longer wavelengths are due to trio and quartet hydrogens, implying that it will be very difficult to separate these two types of hydrogens. In addition, a comparison of the two C$_{102}$H$_{26}$ species shows that the trio and even the duo band intensities can vary greatly depending on the specific structure of the PAH.

\section{Astrophysical Implications}
\label{astro}

The spectroscopic properties of the large irregular PAHs presented in Sect.~\ref{model} are now combined with earlier studies and applied to astronomical observations.  
Before using the fundamental IR spectroscopic properties of PAHs to analyze the major astronomical emission features, it is important to note that they arise from highly vibrationally excited PAHs whereas DFT computations provide harmonic frequencies of vibrationally relaxed molecules, essentially at 0 K.  To compare IR emission spectra with these fundamental frequencies, one must take two aspects of the molecular physics of the emission process into account.  First, as discussed in Sect.~\ref{model}, the transitions must be given bandwidths that are consistent with the natural linewidths of the emission process.  Second, since a small redshift is intrinsic to the emission process, a similarly sized redshift must be applied to the theoretically computed and experimentally measured absorption spectra.  This redshift arises from a sequence of difference bands \citep[e.g.][]{Herzberg}.  There have been a number of experimental studies which measure this redshift as PAHs are raised above their ground vibrational levels.  Some have measured the shift in peak position of the major absorption bands as a function of PAH sample temperature \citep{Flickinger:T:90, Flickinger:91, Colangeli:T:92, Joblin:T:95} while others have measured the emission from highly vibrationally excited, individual PAH molecules in the gas phase \citep{Cherchneff:pah+uvlaser:89, Brenner:benz+naph:92,WilliamsLeone:95, Cook:excitedpahs:98}, mimicking the astronomical situation.

In the astronomical environments, the emission originates from a collection of vibrationally excited PAHs.  Each species is at a different level of excitation, depending on the energy of the absorbed photon which initially excited the molecule and the time since that absorption when the molecule can relax via emission of IR photons in the fundamental bands.  It is far beyond the scope of this paper to treat this process in any detail and apply these corrections to the spectra.  Rather, we follow what has been standard practice in such analyses and adopt an average and approximate value of 15 cm$^{-1}$ for the redshift that arises from sequence bands.  While there is some uncertainty associated with this approach, it is not large.  For example, part of the gas phase emission studies by \citet{Brenner:benz+naph:92} and \citet{WilliamsLeone:95} included the small bicyclic PAH naphthalene (C$_{10}$H$_8$).  In these studies, C$_{10}$H$_8$ was highly vibrationally excited by 40,800 cm$^{-1}$ photons and emission from the CH stretch feature near 3.3 \mum\, (3000 cm$^{-1}$) measured as a function of time after excitation.  In both cases a redshift of some 25 to 30 cm$^{-1}$ is observed within microseconds.  This quickly collapses as the molecules relax by emitting IR photons, and appears to be on the order of 10 cm$^{-1}$ 38 microseconds after excitation.  \citet{WilliamsLeone:95} discuss the magnitude and origin of this sequence band redshift in detail.  Since naphthalene is at least 5 times smaller than the PAHs thought responsible for the astronomical emission features and the heat capacity goes roughly as 3 times the atom number, adopting a 15 cm$^{-1}$ redshift seems a reasonable approximation.  Clearly, there is some uncertainty associated with applying the same 15 cm$^{-1}$ redshift to all the features, however, given that the FWHM of the astronomical bands is on the order of this shift or larger, as with the assumed bandwidth discussed in Sect.~\ref{model}, these idealized spectra can be useful in better understanding the observed astronomical spectra.

For the purposes of the analyses presented here, we present a size range for the PAHs that are the strongest contributors to the emission in each wavelength range.  These ranges are taken from the work of \citet{Schutte:model:93} who derived a model to describe the PAH emission process from first principles.  This model was developed when PAH mid-IR spectroscopy was essentially unexplored, the molecular physics of the emission process largely untested, and years before ISO and Spitzer were launched.  It is time for a new model, one which includes a modern assessment of the sizes of the species that dominate the emission in each wavelength regime.  This caveat should be kept in mind throughout the following.

\begin{figure*}[t!]
\centering
   \begin{minipage}[c]{0.33\textwidth}
      \centering \includegraphics[width=\textwidth]{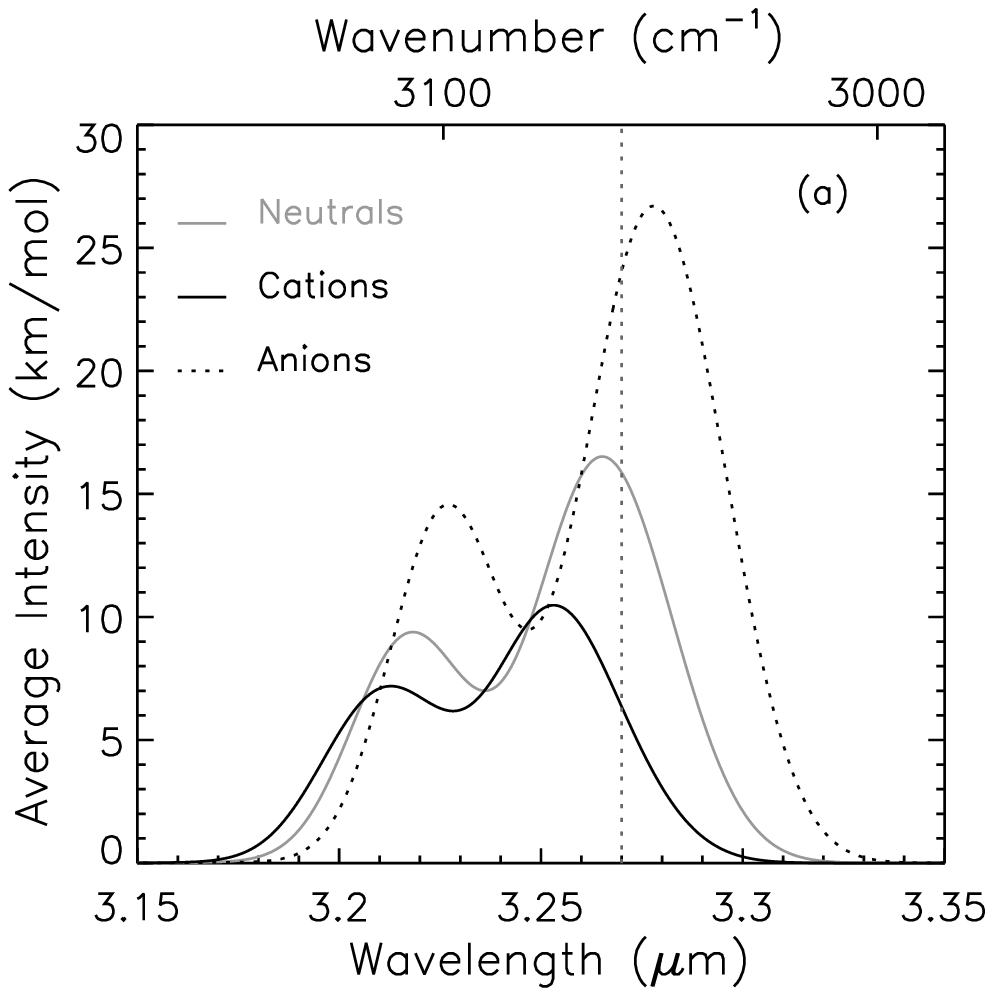}
   \end{minipage}
   \begin{minipage}[c]{0.33\textwidth}
      \centering \includegraphics[width=\textwidth]{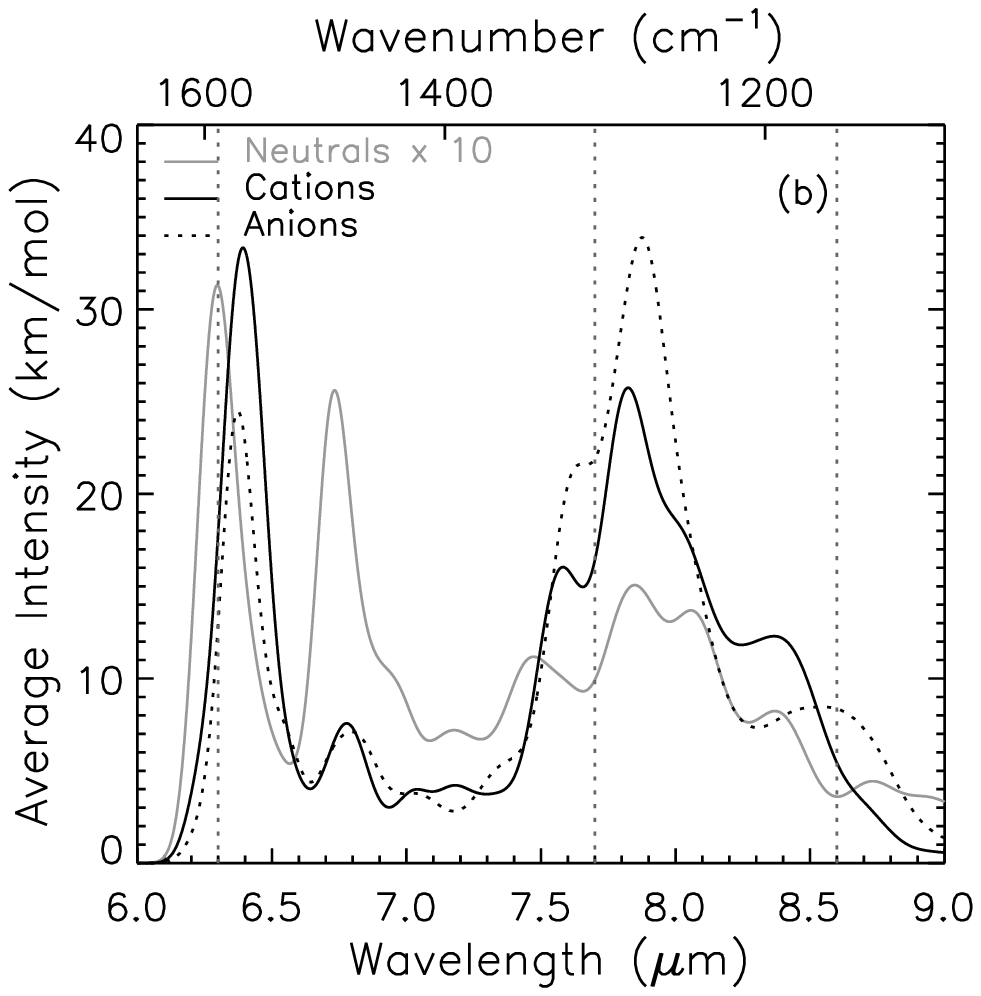}
   \end{minipage}
   \begin{minipage}[c]{0.33\textwidth}
      \centering \includegraphics[width=\textwidth]{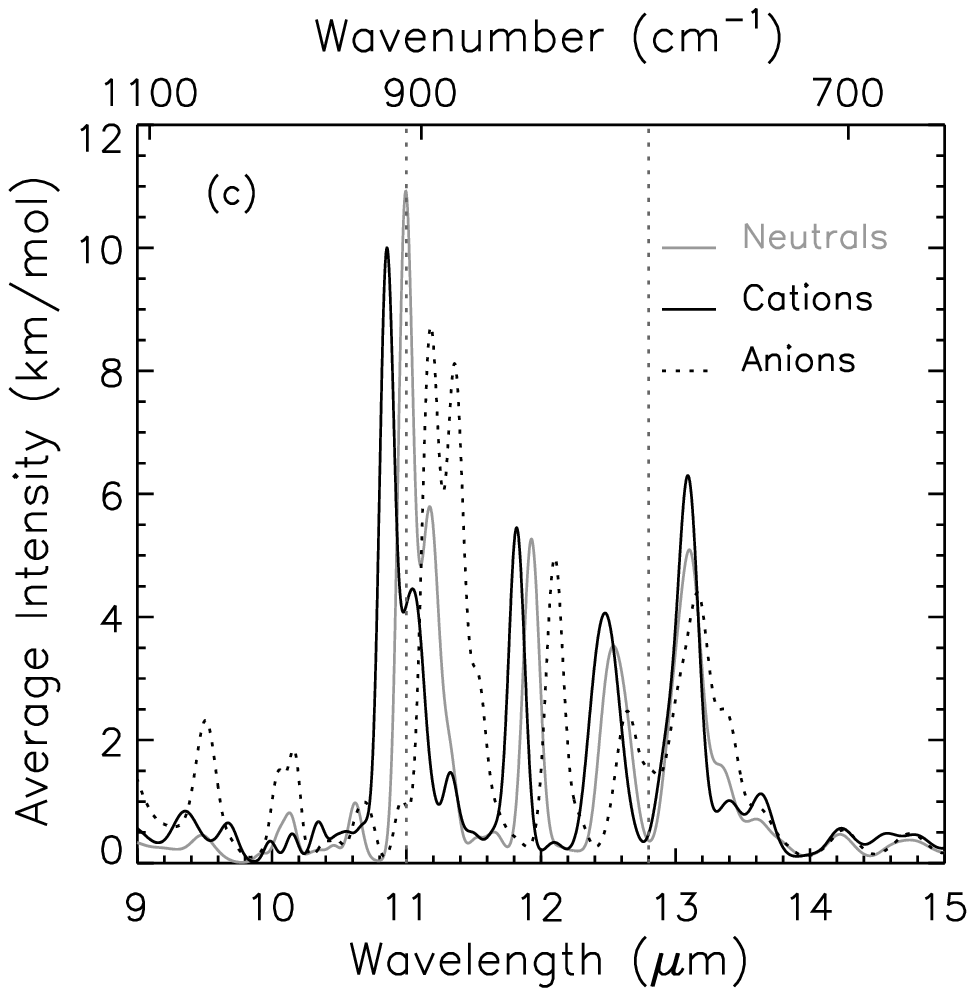}
   \end{minipage}
   \begin{minipage}[c]{0.33\textwidth}
      \centering \includegraphics[width=\textwidth]{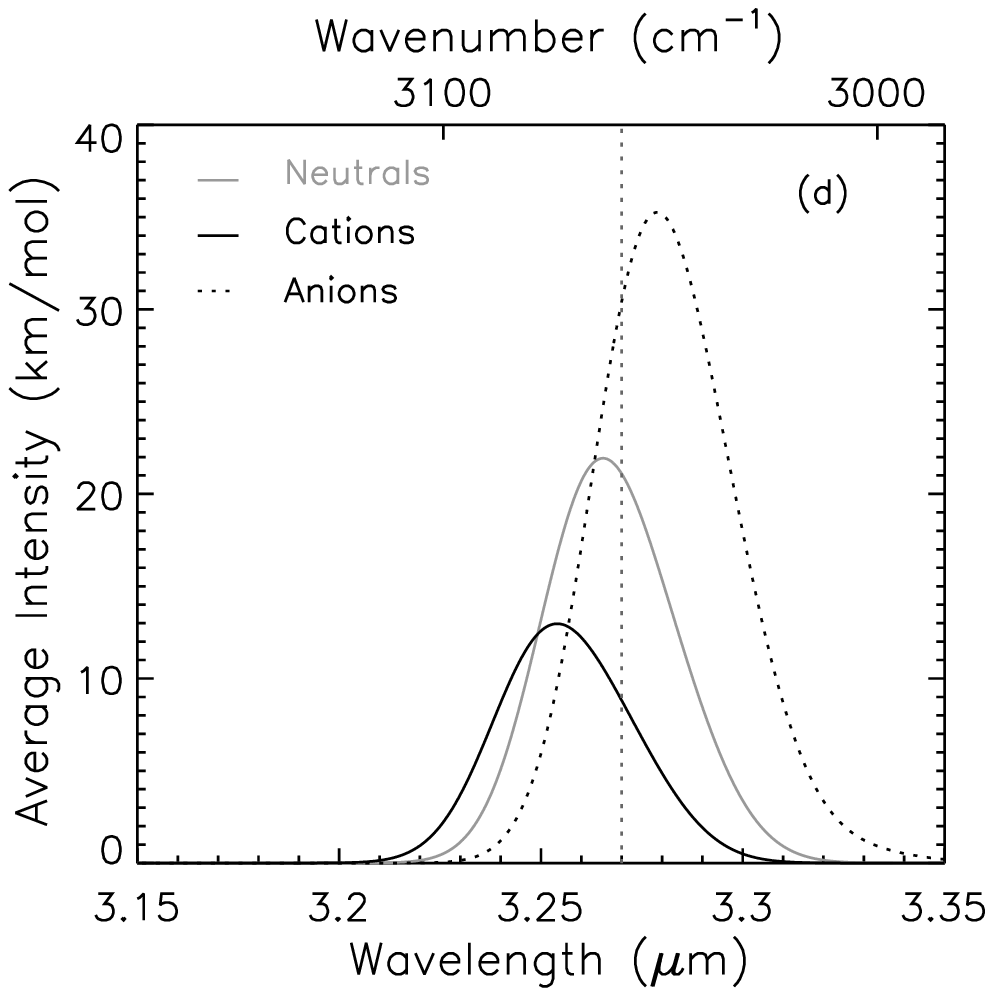}
   \end{minipage}
   \begin{minipage}[c]{0.33\textwidth}
      \centering \includegraphics[width=\textwidth]{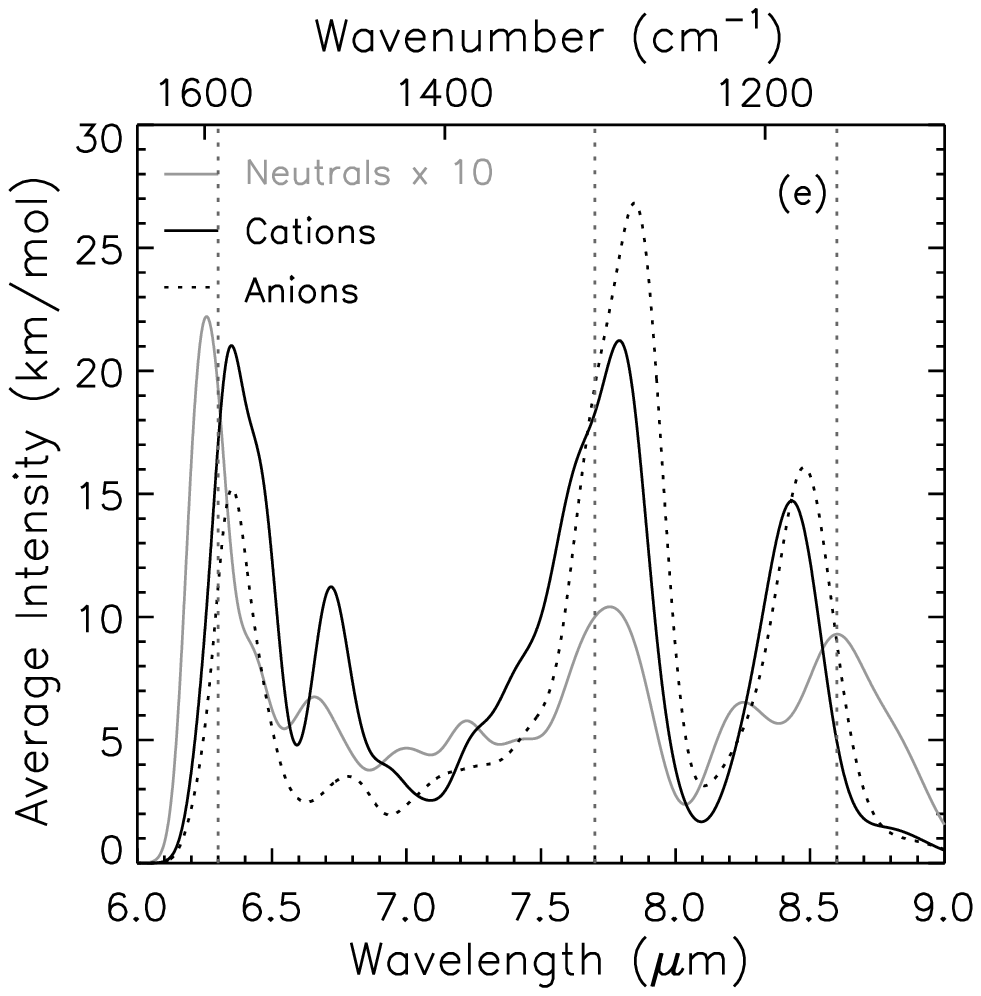}
   \end{minipage}
   \begin{minipage}[c]{0.33\textwidth}
      \centering \includegraphics[width=\textwidth]{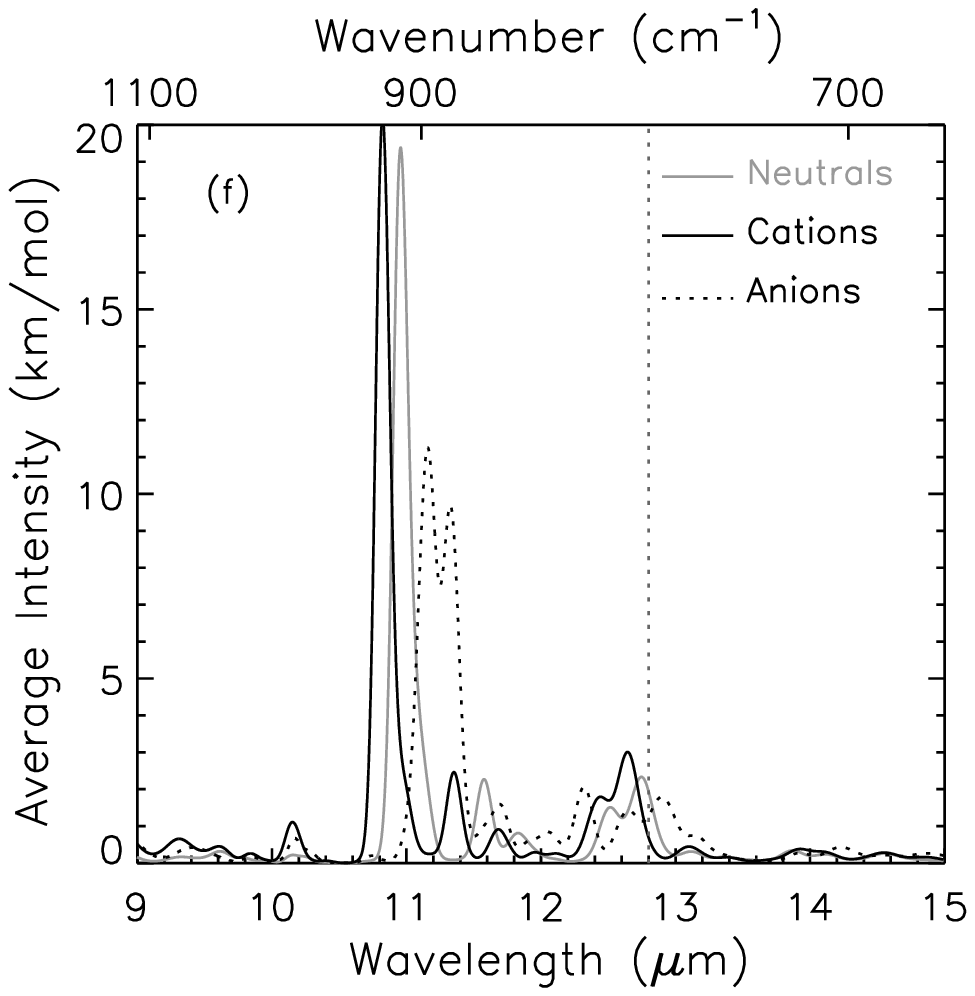}
   \end{minipage}
\caption{The average synthetic absorption spectra for the cation, neutral, and anion forms of the large irregular PAHs considered here (panels a, b, c; \stref\, is not taken into account for the average) compared with the average spectra of the large compact PAH cations, neutrals, and anions treated in Paper 1 (panels d,e,f).  To guide the eye, dotted lines at 3.27, 6.3, 7.7, 8.6, 11.0 and 12.8 $\mu$m\, are also shown.}
\label{fig_average}
\end{figure*}

The top row of Figure~\ref{fig_average} shows the average synthetic spectra of the large irregular PAHs considered here while the bottom row reproduces the average synthetic spectra for the large compact PAH shown in Figure 6 of Paper I.  These data provide a deeper insight into the nature of the astronomical PAH population and place tighter constraints on the molecular structures than previously possible.  The insight gleaned from each vibration specific wavelength region will be discussed in turn, followed by a summary (Sect.~\ref{astro_sum}) in which all constraints are drawn together.  There is a rich literature on the mid-IR spectroscopy of the PAH emission features in each wavelength region discussed below.  A doorway into this literature is provided by the summary reference in the lead paragraph of each section. 

\clearpage
\clearpage

\subsection{The CH stretching vibrations (2.5-3.5 \mum)}
\label{astro_33}

A detailed analysis of the astronomical features produced by the aromatic CH stretching (CH$_{str}$) vibrations was given by \citet{vanDiedenhoven:chvscc:03}. These authors compared the astronomical data with a sample of laboratory and theoretical data comprised of over 100 small neutral and cationic PAHs (C$<$ 40), a few larger PAHs, and 27 small PAH anions.  In Paper I, this analysis was extended to include the spectra of large, symmetric, compact PAHs.  Here we add the spectroscopic properties of the large irregular PAHs presented in Sect.~\ref{model_33}.

\clearpage
\begin{figure}[]
      \centering \includegraphics[width=.5\textwidth]{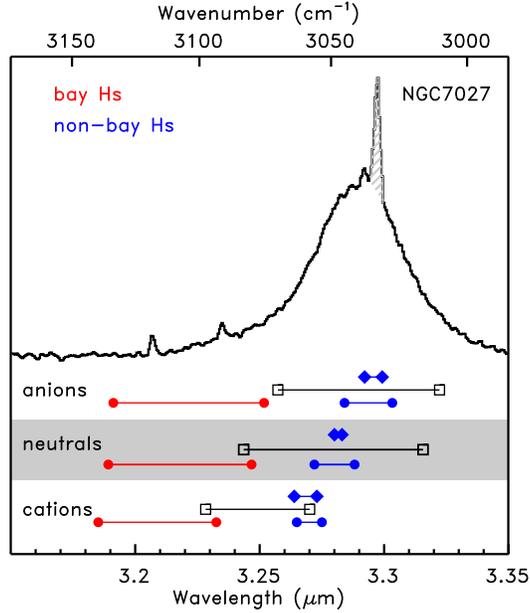}
\caption{Comparison of the astronomical emission feature from NGC\,7027
(class A$_{3.3}$) with the range in peak positions of the CH stretch in various
PAHs.  The striped, narrow grey feature in the spectrum of NGC\,7027 is
the Pf$\delta$ line. The PAH data are redshifted by 15 cm$^{-1}$, a
shift which is intrinsic to the emission process (see discussion in
Sect.~\ref{astro} for details). The connected, filled circles indicate the range
of the CH stretch in the large irregular PAHs considered in this paper
(excluding \stref). Note that the bay and non-bay CH stretch
positions (in red and blue respectively) are represented separately for the large irregular PAHs. The connected, filled diamonds
indicate the range of the CH stretch in the large compact PAHs
§considered in Paper I. The CH stretching modes for the smaller PAH
sample described in Sect.~\ref{astro_33} is represented
by the connected open squares. Here, no distinction has been made between bay and non-bay Hs in the spectra of small PAHs.}
\label{fig_CHstr}
\end{figure}
\clearpage

The differences in the CH$_{str}$ features of irregular PAHs versus those of compact PAHs (shown in Figure~\ref{fig_average}a and \ref{fig_average}d) are striking.  In spite of the similarity in the behavior of the bands arising from the non-bay CH stretches in the irregular PAHs to those in the compact symmetric PAHs, the prominence of the CH$_{str}$ bands associated with bay regions alters the entire appearance of the spectrum.  Figure~\ref{fig_CHstr} compares the 3.29 \mum\, astronomical emission feature from NGC\,7027 with the range in the CH$_{str}$ peak positions for the cation, neutral, and anion forms of the PAHs in the three different samples.  For the smaller PAH sample \citep{vanDiedenhoven:chvscc:03}, only the peak position of the strongest emission band is considered, irrespective of its origin in non-bay or bay Hs. It is clear that the peak positions for the large compact PAHs (filled diamonds) fall in a much narrower region than do the peaks produced by the handful of large irregular PAHs considered here (filled circles) or the sample comprised of small PAHs (open squares).  Upon recognition of the influence bay regions have on the CH$_{str}$ peak positions in large irregular PAHs, we reviewed the experimental spectra in the sample of the small PAHs.  As with the large irregular PAHs, the CH$_{str}$ in small PAHs depends on the presence or absence of bay regions.  For example, the dominant feature in the CH$_{str}$ region of the small, three-ring PAH phenanthrene (C$_{14}$H$_{10}$), which has a bay region, is broad and structured with a FWHM of about 60 cm$^{-1}$. In contrast, the main band in the CH$_{str}$ region of the three ring PAH anthracene, which does not have a bay region and is positioned at similar wavelengths, is much simpler, and has a FWHM of about 13 cm$^{-1}$.  Likewise the feature in the bay region containing four ring PAH benzanthracene (C$_{18}$H$_{12}$) has a FWHM between 50 and 55 cm$^{-1}$ while that in the spectrum of the non-bay region containing 4 ring PAH Pyrene, centered at similar wavelengths, has a FWHM of less than 20 cm$^{-1}$. Likewise, the theoretical data of the smaller PAH sample show the same dichotomy between bay and non-bay Hs as seen in the large irregular PAHs.

The data listed in Table~\ref{t1} and the information conveyed in Figures~\ref{fig_average} and \ref{fig_CHstr} place several important constraints on the astronomical PAH population.  {\it First}, these data show that, while the bulk of the astronomical 3.28 \mum\, band is carried by neutral and negatively charged PAHs, cations can also contribute significant intensity to this band. One can roughly estimate the relative contributions of each charge state by assuming that all emission observed at the 
average peak wavelength for non-bay CH stretching emission of PAHs in a given charge state is only emitted by PAHs of this given charge state. For example, the average peak wavelength for non-bay CH stretching emission for the cations presented in this paper is at $\sim$3.25 \mum. The cationic contribution can then be estimated by the ratio of the intensity observed at this wavelength (which takes a redshift of 15 cm$^{-1}$ into account) in the continuum-subtracted spectrum of, for example, NGC~7027 to the average intrinsic intensity per CH stretch (I(CH)) for non-bay hydrogen atoms for these cations. The average intensity per CH stretch (I(CH)) for non-bay hydrogen atoms in the PAH cations listed in Table~\ref{t1} is 26.5km/mol. By determining the ratio of the CH$_{str}$ for the neutral and cations, one can estimate the contribution of each charge state to the 3.3 \mum\, band.  For the spectrum of NGC~7027 shown in Fig.~\ref{fig_CHstr}, the cation to neutral ratio is then estimated to be $\sim$0.90. Given that the observed differences in the 3.3 \mum\, PAH profile are relatively small \citep{vanDiedenhoven:chvscc:03}, this ratio holds for most astronomical objects. Thus, nearly equal amounts of CH$_{str}$ modes from cations and neutral PAHs can contribute to the 3.3 \mum\, band. Given that the absolute intensity per CH stretch for neutral PAHs is about twice that of the cations, neutral PAHs will dominate the total flux contributed by this two species. Similarly, the anion to neutral ratio is estimated to be $\sim$ 0.70. This modifies the earlier conclusion in \citet{vanDiedenhoven:chvscc:03} and paper I that cations could be excluded as a major contributor to this feature, and argues against an origin of the 3.28 \mum\, band exclusively in neutral PAHs as commonly claimed in the astrophysics literature, a conclusion based on correlations with the 11.2 \mum\, band and earlier laboratory data of small PAHs. If the redshift due to the difference bands is 30 cm$^{-1}$ instead of 15 cm$^{-1}$, the cation to neutral ratio is 1.5 and the anion to neutral ratio is 0.5. Hence these ratios clearly depend on the assumed redshift. However, this does not change the conclusions. {\it Second}, as proposed in Paper I, the small separation between the prominent (non-bay) CH$_{str}$ peak positions for the neutral and anion compact PAH forms suggests that the Type 1 and Type 2 band profiles identified by \citet{Tokunaga:33prof:91} and referred to as the A$_{3.3}$ and B$_{3.3}$ band types by \citet{vanDiedenhoven:chvscc:03} may reflect the relative amounts of neutral/cationic and anionic PAHs producing the blended feature.  {\it Third}, since the smallest members of the astronomical PAH population dominate the emission in the 3 \mum\, region \citep{Schutte:model:93, ATB} and hydrogen atoms spanning bay regions produce a distinct feature peaking between about 3.21 and 3.23 \mum, one can place an upper limit on the relative number of bay to non-bay hydrogens in these PAHs thereby constraining the overall structures of the smallest astronomical PAHs.  Disregarding the values for \stb, the very unusual species for which the bay hydrogens are internal to the hexagonal carbon atom network (Figure 1), the average intensity per CH stretch (I(CH)) for bay hydrogen atoms listed in Table~\ref{t1} are 24.5, 32.2, and 52.3 km/mol for the cations, neutrals, and anions respectively.  The corresponding average  intensities per CH stretch for the non-bay hydrogen atoms in the irregular PAHs are 21.6, 34.6, and 54.5 km/mol and those for the comparably sized compact PAHs considered in Paper 1 are 26.5, 40.7, and 64.4 km/mol.  This shows the intrinsic intensities of the bay and non-bay CH stretching bands are roughly the same. Thus, the ratio of non-bay hydrogens to bay hydrogens in the astronomical PAH population can be estimated from the ratio of the peak intensity of the continuum subtracted 3.3 \mum\, feature to the residual above this continuum at 3.23 and 3.24 \mum\, respectively for neutrals and anions (applying a 15 cm$^{-1}$ redshift). From the spectrum of NGC\,7027 shown in Figure~\ref{fig_CHstr}, the bay/non-bay hydrogen atom ratios for the anion and neutral cases are about 0.12 and 0.09 respectively. 
Given that the observed differences in the 3.3 \mum\, PAH profile are relatively small \citep{vanDiedenhoven:chvscc:03}, this ratio holds for most astronomical objects.  Hence, these low ratios imply that $\sim$90\% of the PAH emission in the 3 \mum\, region comes from PAHs with non-bay, aromatic edge structures. This value decreases to $\sim$75\% if a 30 cm$^{-1}$ redshift is taken. {\it This observation forces one to conclude that the vast majority of astronomical PAHs responsible for the emission in the 3 \mum\, region are compact - not irregular, elongated or bent.} Given that emission in the 3 \mum\, region samples the smallest members of the PAH population, this requirement for a PAH population dominated by compact structures must hold for the very smallest members of the population.  This limitation of bay regions to less than 10\% of the edge structures on the smallest members of the astronomical PAH family limits the number of species responsible for the 3.3 \mum\, feature to a handful, a point considered further in Sect.~\ref{astro_sum}.

\subsection{The CC stretching and CH in-plane bending vibrations (5-9 \mum)}
\label{astro_62}

The astronomical emission bands arising from PAH CC stretching (CC$_{str}$) and CH in-plane (CH$_{ip}$) bending vibrations were discussed in great detail and analyzed with a dataset similar to that described in Sect. 3.1 by \citet{Peeters:prof6:02}.  In Paper 1, this work was expanded to include large, symmetric, compact PAHs.  Here this is extended to incorporate the spectroscopic properties of the large, irregular PAHs described in Sect.~\ref{model_62}.

\begin{figure}[]
      \centering \includegraphics[width=.5\textwidth]{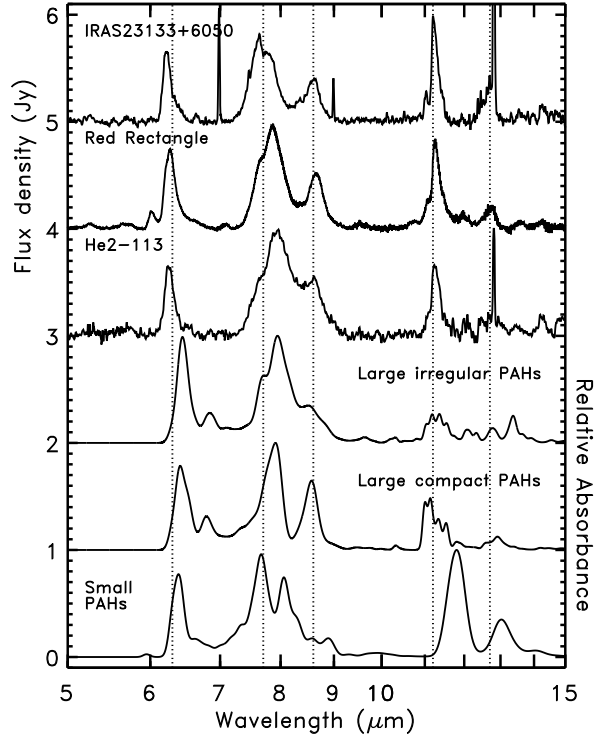}
\caption{Astronomical PAH emission spectra from 5 to 15 $\mu$m
  compared with the spectra of PAH mixtures. Shown are the scaled, continuum
  subtracted, ISO-SWS spectrum of IRAS\,23133+6050 representing class A
  PAH profiles {\bf (top)}; the scaled continuum subtracted ISO-SWS spectrum
  of the Red Rectangle, representing class B PAH profiles {\bf
  (second)}; the scaled continuum subtracted ISO-SWS spectrum of the PN He2-113, representing slightly red-shifted class B PAH profiles to illustrate possible variations within class B in this wavelength region {\bf (third)}; the average spectrum of the large irregular PAH cations,
  neutrals and anions considered here {\bf (fourth)}; the average
  spectrum of the large compact PAH cations, neutrals and anions from
  paper I {\bf (fifth)}; and the composite absorption spectrum
  generated by co-adding the individual spectra of 11 PAHs, reproduced
  from \citet[][see text for composition, {\bf bottom}]{Peeters:prof6:02}.  A redshift of 15~cm$^{-1}$, intrinsic to the emission
  process, has been applied to the positions of the large and small
  PAH spectra (see discussion in Paper I for details). To guide the
  eye, dotted vertical lines are also shown at the positions of the
  major astronomical bands at 6.3, 7.7, 8.6, 11.2 and 12.7 \mum.}
\label{fig_6to9}
\end{figure}

Figure~\ref{fig_average}b shows the average spectra of the different charge forms of the irregular PAHs in the 5 to 9 \mum\, region and Figure~\ref{fig_average}e shows the corresponding spectra for the large compact PAHs presented in Paper I.  Overall, the spectra in the CC stretching and CH in-plane bending region resemble one another, in sharp contrast with the striking differences between the bands in the CH$_{str}$ region (Figs.~\ref{fig_average}a and \ref{fig_average}d) and CH$_{oop}$ region (Figs.~\ref{fig_average}c and \ref{fig_average}f).  Figure~\ref{fig_6to9} compares the emission spectra from the HII region IRAS\,23133+6050, the post-AGB star the Red Rectangle and the [WC10] PN He2-113 as studied by \citet{Peeters:prof6:02} to three different collections of PAH spectra: the average spectrum produced by the large irregular PAHs computed here, the average spectrum of the large compact PAHs described in Paper I, and an earlier `best fit' average mixture of small PAHs\footnote{The composite spectrum of 11 PAHs
consists of 22\% neutral coronene (C$_{24}$H$_{12}$); 19\%
3,4;5,6;10,11;12,13-tetrabenzoperopyrene cation (C$_{36}$H$_{16}^+$);
15\% coronene cation (C$_{24}$H$_{12}^+$); 7\% dicoronylene cation
(C$_{48}$H$_{20}^+$); 7\% benzo[b]fluoranthene cation
(C$_{20}$H$_{12}^+$); 7\% benzo[k]fl uoranthene cation
(C$_{20}$H$_{12}^+$); 7\% neutral naphthalene (C$_{10}$H$_8$); 4\%
naphthalene cation (C$_{10}$H$_{8}^+$); 4\% phenanthrene cation
(C$_{14}$H$_{10}^+$); 4\% chrysene cation (C$_{18}$H$_{12}^+$); 4\%
tetracene cation (C$_{18}$H$_{12}^+$).}$^,$.

The prominent bands near 6.3 \mum\, in the spectra shown in Figs.~\ref{fig_average}b and \ref{fig_average}e originate in CC stretching vibrations.  As shown in the figure, the behavior of these bands in the spectra of the irregular PAHs is similar to that for the compact PAHs.  In both cases, the anion and cation bands overlap, with the anion feature a little weaker than and slightly to the blue of the cation band.  The CC$_{str}$ peak falls close to 6.4 \mum\, in all of these large irregular PAHs, behavior that is consistent with all previous work on PAH spectroscopy.  However, as can be seen in Figure~\ref{fig_6to9}, this cannot account for the peak position of the Class A 6.2 \mum\, astronomical band. This inability of the CC$_{str}$ in pure PAHs to match the Class A band has lead to the suggestion that astronomical PAHs contain internal nitrogen atoms \citep[PANHs,][]{Peeters:prof6:02, Hudgins:05}.  The new data on irregular PAHs reinforces that suggestion.

As with the compact PAHs considered in Paper I, some irregular PAH cations and anions also have a minor feature between 6.7 and 6.8 \mum.  A weak interstellar feature has been detected close to this position near 6.8 to 6.9 \mum\, and has generally been attributed to CH deformations in aliphatic side groups on PAHs.  Figures~\ref{fig_average}b and \ref{fig_average}e show that PAHs without aliphatic components have bands in this region, removing the necessity to invoke these groups to account for this band. 
As discussed in Sect.~\ref{model_62}, while this band is strong in the spectrum of some neutral PAHs, it can be neglected since the spectra of the cations and anions are so much stronger.  

\clearpage
\clearpage

Moving now to the prominent bands peaking near 7.7 \mum, one is struck by the somewhat similar profiles and close overlap between the features produced by each charge state of the irregular PAHs (Figure~\ref{fig_average}b).  As is the case for the large compact PAHs, the average anion peaks about 0.1 \mum\, to the red of the average cation peak.  This does not hold for the 6.2 \mum\, region, where the anions are blue-shifted compared to the cations by a very small amount. The overall impression is that the main feature in all charge states of the irregular PAH spectra is broader than that for the compact PAHs.  However, this is largely due to the lack of a sharp drop on the long wavelength side of the band in the spectra of the irregular PAHs.  The CH$_{ip}$ bending vibrations which are the modes responsible for the distinct 8.6 \mum\, feature in the spectra of compact PAH cations and anions (Figure~\ref{fig_average}e) produce bands that are weaker than their counterparts in the compact PAH spectra.  These bands are also blue-shifted by one to two tenths of a micron in the irregular PAHs and blend with the strong bands that form the 7.8 \mum\, feature, producing the long wavelength wing.  This shifting occurs because the irregular edge structures of these PAHs impose variation in the vibrational force field along the edge of the molecules.  Given that the modes which produce bands longward of 8 \mum\, involve a CH vibration, slight shifting of these band's fundamental frequencies in any given molecule with different edge structures is expected.  The absence of a distinct band near 8.6 \mum\, due to the CH$_{ip}$ bend in these irregular PAHs is consistent with the conclusions drawn from the CH$_{str}$ band discussed in Sect.~\ref{astro_33}, that compact PAH structures are required to account for the distinct astronomical feature observed in both regions.  Some edge irregularities can be invoked to account for the spectra from those objects which do not have a prominent 8.6 \mum\, band on the long wavelength wing of the 7.7 \mum\, feature, but these must be associated with somewhat larger PAHs than those which contribute most to the emission at 3 \mum.

Perusal of Figure~\ref{fig_6to9} shows that the astronomical spectra from 5 to 9 \mum\, may be accommodated by emission from PAH populations that includes both small and large PAHs.  This is discussed further in Sect.~\ref{astro_sum} where the constraints from the other spectral regions are also considered.

\subsection{The CH out-of-plane bending vibrations (9-15 \mum)}
\label{astro_112}

A careful analysis of the astronomical features produced by the aromatic CH$_{oop}$ bending vibrations was given by \citet{Hony:oops:01}.  This was based on the experimental study of 20 PAHs (C$_{10}$H$_8$ to C$_{32}$H$_{14}$) which sampled the different types of edge structures and hydrogen adjacency classes possible \citep{Hudgins:tracesionezedpahs:99}. This was expanded to incorporate the spectroscopic properties of large compact PAHs in Paper I and here we include the properties of the large irregular PAHs presented in Sect.~\ref{model_112}.  The PAH features in the CH$_{oop}$ region originate in solo, duo, trio, and quartet hydrogens.  As shown in Sect.~\ref{model_112}, with the exception of the duo hydrogen bands, the regions in which the CH$_{oop}$ vibrations fall for the different types of adjacent hydrogen atoms in large irregular PAHs are consistent with those in earlier studies.

\clearpage
\begin{figure}[t!]
      \centering \includegraphics[width=.45\textwidth]{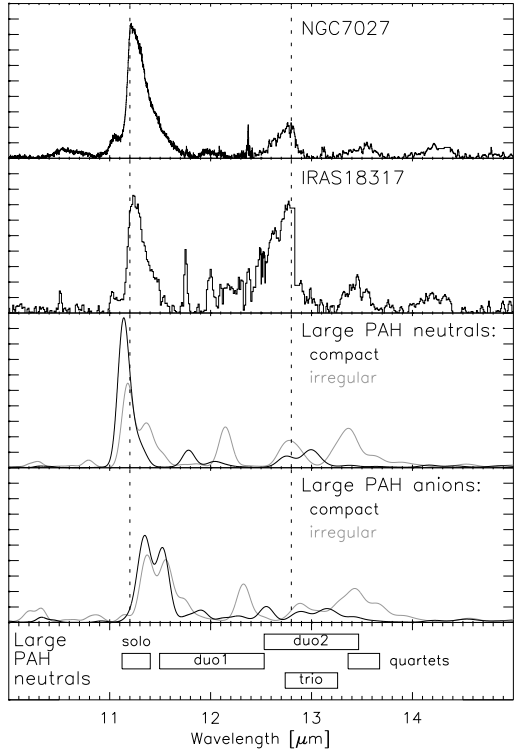}
\caption{Astronomical PAH emission spectra from 10 to 15 $\mu$m
   compared with the average spectrum of very large, compact
    PAHs (paper I) and the average spectrum of very large
   irregular PAHs (this paper) for both the neutral and anionic state. Shown are 
   the continuum subtracted IR
   spectrum of the planetary nebula NGC\,7027 {\bf (top)}; the
   continuum subtracted IR spectrum of the \HII \,region IRAS\,18317-0757
   {\bf (second)}; the average spectrum of the neutral large PAHs
  considered in this paper and paper I {\bf (third)}; the average spectrum of the large PAH anions
  considered in this paper and paper I {\bf (fourth)}; and updated ranges for the CH$_{oop}$ bending modes based on the spectra of large PAHs presented here and in Paper I {\bf (bottom)}. The data in the lower three
   frames have been redshifted by 15 cm$^{-1}$ to correct for the
   wavelength shift between the absorption and emission process (see
   discussion in Sect.~\ref{astro} for details). Note that for the astronomical sources, a [NeII] fine-structure line emission was removed by fitting a Gaussian profile to the line with a FWHM corresponding to the instrumental resolution. }
\label{fig_oops}
\end{figure}
\clearpage

Figure~\ref{fig_average}c shows the average spectra of the different charge forms of the large irregular PAHs in the 9 to 15 \mum\, region and figure~\ref{fig_average}f shows the corresponding spectra for the large compact PAHs presented in Paper I.  The large spectral differences between the spectrum in the CH$_{oop}$ region associated with compact PAHs versus those associated with irregular PAHs are both striking and expected.  All the species in the compact PAH sample contain only solo and duo hydrogens whereas the irregular PAH sample contains solo, duo, trio, and quartet hydrogens.  Figure~\ref{fig_oops} compares the emission spectra from the planetary nebula NGC\,7027 and the HII\, region IRAS\,18317-0757 to the average spectrum of irregular and compact neutral PAHs, and the average spectrum of irregular and compact PAH anions.  The bottom frame of Figure~\ref{fig_oops} shows the CH$_{oop}$ regions characteristic of solo, duo, trio, and quartet hydrogens that are based on the spectra of the large PAHs considered here and in Paper I. These updated values supersede those shown in Fig.~\ref{fig_112} which are appropriate for PAHs comprised of fewer than 32 C atoms.

The spectra in Figure~\ref{fig_oops} show the extremes in the 11.2/12.7 PAH band intensity ratios reported by \citet{Hony:oops:01}.  The weak 11.0 \mum\, astronomical band evident in the spectra of both objects is not reproduced in the PAH data since these spectra are of neutral and negatively charged species and the astronomical band has been attributed to positively charged PAHs \citep{Hudgins:tracesionezedpahs:99}.  The bands at 13.5 and 14.3 \mum\, in the astronomical data are assigned to CH$_{oop}$ bends that arise from quartet and quintet hydrogens respectively \citep{Hony:oops:01}.  While these astronomical bands are absent in the compact PAH spectra because they contain only solo and duo hydrogens, a counterpart to the 13.5 \mum\, astronomical band is evident in the average irregular PAH spectrum because these do contain quartet hydrogens.  However, since they don't contain quintet hydrogens, their spectra do not have a feature near 14.3 \mum.

\clearpage
\begin{table}
\caption{\label{t4} Representative band strengths (km/mol) for different hydrogen adjacency classes in PAHs. See Section~\ref{astro_112} for details.}
\begin{center}
\begin{tabular}{lcccccc}
\hline \\[-5pt]
Large PAHs & \multicolumn{2}{c}{Cation} &\multicolumn{2}{c}{Neutral} &\multicolumn{2}{c}{Anion}\\[5pt]

& \multispan6 \hfil {\bf solo} \hfil \\
irregular\tablenotemark{a} & \multicolumn{2}{c}{19.1} & \multicolumn{2}{c}{19.4} & \multicolumn{2}{c}{23} \\
compact\tablenotemark{b} & \multicolumn{2}{c}{24.9} & \multicolumn{2}{c}{25.5} & \multicolumn{2}{c}{27.0} \\[5pt]

& \multispan6 \hfil {\bf duo} \hfil \\
& duo1 & duo2 & duo1 & duo2 & duo1 & duo2 \\
irregular\tablenotemark{a} & 3.8 & 11.4 & 4.4 & 10.2 & 3.8 & 9.2\\
compact\tablenotemark{b} & 5.7 & 2.6 & 4.2 & 3.2 & 3.7 & 5.1 \\[5pt]

& \multispan6 \hfil {\bf trio} \hfil \\
irregular\tablenotemark{a} & \multicolumn{2}{c}{5.4}& \multicolumn{2}{c}{5.4}& \multicolumn{2}{c}{8.6}\\
compact\tablenotemark{b} & \multicolumn{2}{c}{no trios}& \multicolumn{2}{c}{no trios}& \multicolumn{2}{c}{no trios}\\[5pt]

& \multispan6 \hfil {\bf quartet} \hfil \\
irregular\tablenotemark{a} & \multicolumn{2}{c}{9.6}& \multicolumn{2}{c}{9.1}& \multicolumn{2}{c}{13.1}\\
compact\tablenotemark{b} & \multicolumn{2}{c}{no quartets}& \multicolumn{2}{c}{no quartets}& \multicolumn{2}{c}{no quartets}\\[5pt]

\hline \\[-10pt]
\end{tabular}
\end{center}
\noindent
\tablenotemark{a}{Values determined by dividing the intensities of the band corresponding
to the appropriate hydrogen adjacency class listed in Table~\ref{t3} by the number of hydrogen atoms in the corresponding class. Then, taking the average of these for all large irregular PAHs.}
\tablenotemark{b}{Values determined by averaging the per CH intensities listed in Table 4,
Paper I, for the 5 PAHs ranging in size from C$_{78}$H$_{22}$ to C$_{130}$H$_{28}$.}
\end{table}
\clearpage

As described in \citet{Hony:oops:01}, the intensities of the astronomical CH$_{oop}$ emission bands corresponding to each hydrogen adjacency class can be used to estimate the relative amounts of these edge structures in the astronomical PAH population.  Here, the earlier conclusions of \citet{Hony:oops:01} are reevaluated based on the new information contained in the spectra of compact PAHs from Paper I and the irregular PAHs considered here.  The average intensity of each CH$_{oop}$ mode, per CH, for each hydrogen adjacency class has been determined for the irregular PAHs.  These are listed in Table~\ref{t4} where they are compared with the corresponding values for the large compact PAHs from Paper I.  We restrict the remainder of this analysis to the spectra of the large PAHs presented here and in Paper I since these are comparable in size to those which dominate the astronomical emission in this wavelength range. The behavior of the intensities corresponding to each adjacency class can be summarized as follows.

\paragraph{Solo hydrogens} Two conclusions can be drawn from the intensities of solo hydrogens listed in Table~\ref{t4}.  First, solo CH$_{oop}$ band intensities are two to ten times greater than those for the other types of hydrogen adjacency classes.  Second, an integrated band intensity of about 22 km/mol seems to hold for all large PAHs studied to date regardless of charge state or size.

\paragraph{Duo hydrogens} The situation regarding the duo hydrogens is not as straightforward.  As discussed in Paper I, for PAHs with more than duo hydrogens, the duo band splits into two components, labeled duo1 and duo2 in Paper I, and listed as such here in Table~\ref{t4}.   This is the case for most PAHs.  The integrated intensities of the duo1 bands in Table~\ref{t4} are all close to 4 km/mol, regardless of charge, structure, or size.  This band is 5 to 6 times weaker than the solo band and generally falls between about 11.3 and 12.3 \mum. In contrast, the integrated intensities of the duo2 bands differ between factors of about 2 to 5 between the large compact and irregular PAHs.  Given the entries in the table, a representative value for the duo2 band of about 5 km/mol seems reasonable. The duo2 band falls between about 12.3 and 12.8 \mum\, in the large irregular PAHs considered here and between 12.5 and 13.2 \mum\ in the compact PAHs treated in Paper I. 

\paragraph{Trio hydrogens} The integrated band intensities for the trio modes in neutral and positively charged irregular PAHs are about 5 km/mol, a value again some 4 to 5 times weaker than that for the solo band.  The anion band strength is higher.  The trio band falls between 12.9 and 13 \mum\, in the different charge states of the two PAHs with trio hydrogens in the large, irregular  PAH sample.  

\paragraph{Quartet hydrogens} The integrated band strengths for the quartet modes in the large irregular PAHs considered here do not show much variance.  The entries in Table~\ref{t4} suggest that an integrated band intensity of about 10 km/mol can be taken as representative for quartet hydrogens.  The quartet hydrogen band falls between 13.1 and 13.4 \mum\, in the large irregular PAHs. 

The representative hydrogen adjacency class band strengths discussed above are summarized in Table~\ref{t5}.\\

\clearpage
\begin{table}
\caption{\label{t5} Representative band strengths (km/mol) for different hydrogen adjacency classes in neutral PAHs. See Section~\ref{astro_112} for details.}
\begin{center}
\begin{tabular}{ccccc}
\hline \\[-5pt]
Solo & \multicolumn{2}{c}{Duo} & Trio & Quartet \\
22 & 4 (1)\tablenotemark{a} & 5 (2)\tablenotemark{a} & 5 & 10 \\[5pt]
\hline \\[-10pt]
\end{tabular}
\tablenotetext{a}{(1) and (2) indicate duo1 and duo2 band values.}
\end{center}
\noindent
\end{table}
\clearpage

As described in \citet{Hony:oops:01}, Sect. 5, one can derive the types and sizes of interstellar PAH structures in the emission zones by analyzing the peak positions and integrated band strengths of the astronomical features with theoretical and laboratory data such as given in Tables~\ref{t4} and \ref{t5}.  Summarizing, this is accomplished as follows.  From Table~\ref{t5}, one obtains the per hydrogen band intensity ratios of the solo-to-duo1, solo-to-trio, and solo-to-quartet hydrogen bands of 5.5, 4.4, and 2.2 respectively.  \citet{Hony:oops:01} derived similar solo-to-duo (6.2) and solo-to-quartet (2.1) ratios, but a nearly 50\% lower solo-to-trio ratio (2.6) because the overlap of the duo2 band with the trio band was not known at the time.  Next, the corresponding ratios for the astronomical bands are derived from the observations.  The observed intensity ratios are then converted to the relative number of hydrogen atoms in each adjacency class with the data in Tables~\ref{t4} and ~\ref{t5}.  Finally, from the number of hydrogen atoms in each adjacency class, one can derive the number of the different PAH edge structures or {\it groups} that produce the solo, duo, trio, and quartet bands in the astronomical spectra.

\begin{figure*}[]
      \centering \includegraphics[height=.95\textheight]{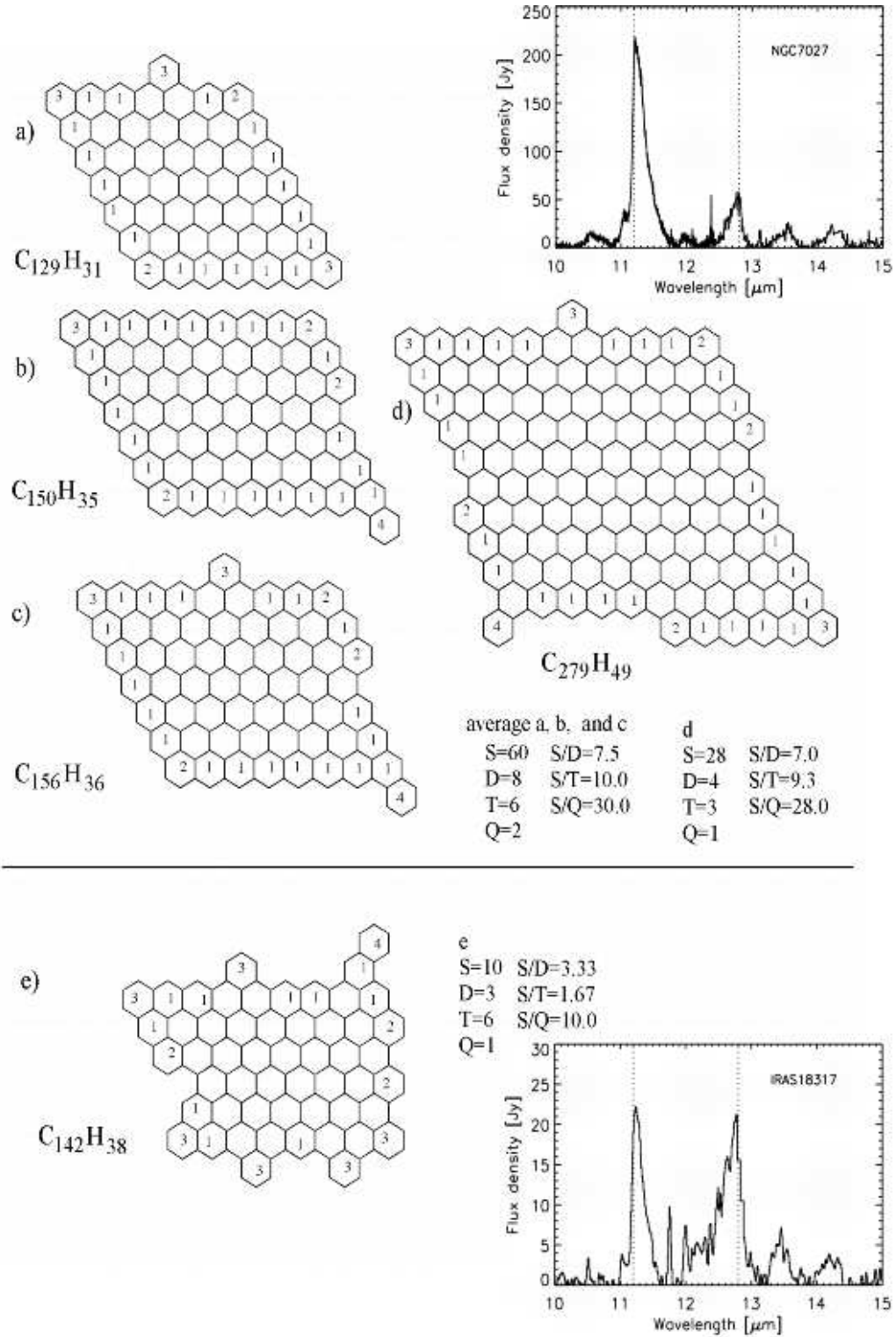}
\caption{Examples of large PAH structures implied by the 10-15 \mum\, spectra of NGC\,7027 {\bf (top)} and IRAS\,18317 {\bf (bottom)}. Two possibilities of structures, one being the combination of three structures (a,b,c),  are shown for NGC\,7027.}
\label{eg_structure}
\end{figure*}

The conclusions in \citet{Hony:oops:01} regarding the relative numbers of different edge structures on astronomical PAHs should be slightly revised.  As described above, the spectroscopic properties of large compact and irregular PAHs show that only about half of the intensity in the 12.7 \mum\, emission band can be attributed to trio hydrogens. Thus the ratios of different edge {\it groups} presented in Table~\ref{t4} of \citet{Hony:oops:01}, should be halved.  Table~\ref{t6} is the updated version of their table which presents the ratios of the various edge {\it groups} on the astronomical PAH populations in NGC\,7027 and IRAS\,18317.  Fig.~\ref{eg_structure} shows example structures that correspond to these ratios. These structures will be considered further in the next section where the constraints from the entire emission spectrum are taken into account.

\clearpage
\clearpage

\begin{table}
\caption{\label{t6} Relative number of solo, duo, trio, and quartet groups in NGC\,7027 and IRAS\,18317-0757. As discussed in Section 3.3, these entries are updated from the values listed in Table 4 from \citet{Hony:oops:01}.}
\begin{center}
\begin{tabular}{cccc}
\hline \\[-5pt]
& Solo/Duo & Solo/Trio & Solo/Quartet \\
NGC\,7027 & 7.7 & 9.2 & 28.5 \\
IRAS\,18317 & 3.4 & 1.6 & 10.2 \\
\hline \\[-10pt]
\end{tabular}
\end{center}
\noindent
\end{table}
\clearpage

\subsection{Tying it all together -- Important implications for the astronomical PAH population}
\label{astro_sum}

\begin{figure*}[]
      \centering \includegraphics[width=.8\textwidth, angle=-180]{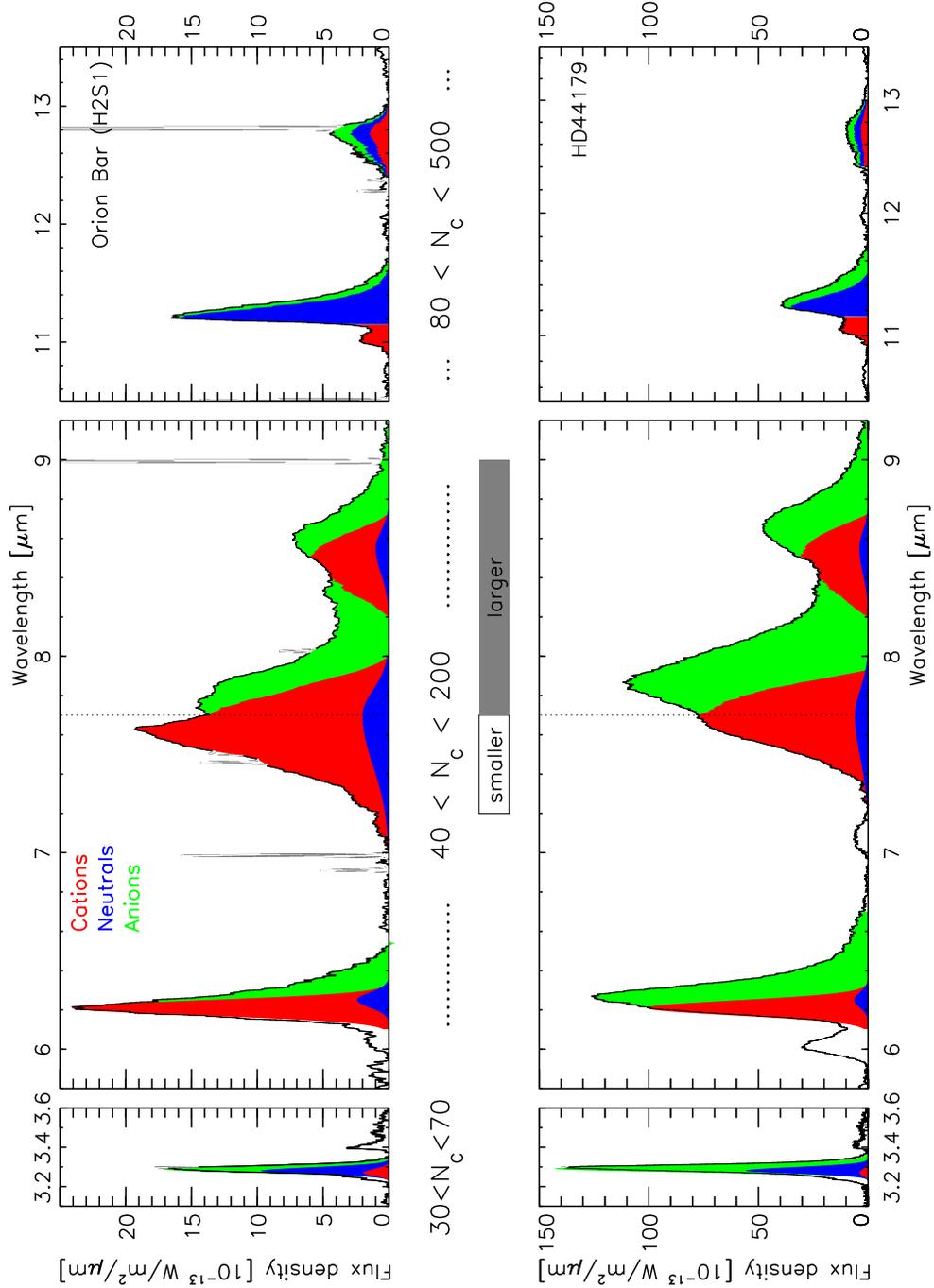}
\caption{A {\it qualitative} representation of the size and charge of the PAHs producing the prominent emission features in the class A and class B objects represented by the Orion Bar (H2S1) and the Red Rectangle (HD44179). The size ranges listed under each region are inferred from Fig. 9 of \citet{Schutte:model:93}. The contributions of different PAH charge states to each feature are careful approximations based on the work presented here and in Paper I. Note that contributions of neutrals to the 6.2 \mum\, band are certainly possible when considering PANHs (see Sect. ~\ref{model_62}). The spectrum of the Orion Bar and the Red Rectangle were obtained with ISO-SWS \citep{Kessler:iso:96, deGraauw:sws:96}.}
\label{fig_final}
\end{figure*}

Here we bring the spectroscopic properties of the dominant PAH features together to place constraints on the overall astronomical PAH population.  The general properties of PAHs responsible for the individual emission features are shown in Fig.~\ref{fig_final}.

\paragraph{The 3.29 \mum\, Feature} We start by considering the species that dominate the small end of the astronomical PAH population, i.e. those species that produce the 3.29 \mum\, band. More than 80\% of the emission of this band, which originates in the CH$_{str}$,  is thought to come from PAHs containing between about 30 to 70 C atoms \citep{Schutte:model:93}. The narrow range in observed peak position coincides with those from neutral, cationic, and anionic PAHs.  Neutral and cationic species likely contribute more to the Type 1 or A3.3 band and anions more to the Type 2 or B3.3 band \citep{Tokunaga:33prof:91,  vanDiedenhoven:chvscc:03}.  Further, our best estimate suggests that less than $\sim$10\% of these PAHs contain bay regions, though this changes to less than $\sim$25\% when a 30 cm$^{-1}$ redshift is applied. Nevertheless, this immediately forces one to conclude that the smallest astronomical PAHs have overall molecular geometries and edge structures that are very similar to the compact, regular PAHs described in Paper I. 

\paragraph{The 5 -- 9 \mum\, Features} Next, we move on to PAHs which are slightly larger, the species responsible for the dominant features in the 5 to 9 \mum\, region. More than half of this emission originates in PAHs containing some 40 to 200 carbon atoms, but larger species comprising several hundred carbon atoms can also contribute to the bands and plateaus in this region  \citep{Schutte:model:93}. Consider first the strong emission band between 6.2 and 6.3 \mum. This prominent CC$_{str}$ feature is particularly revealing about the astronomical PAH population. As noted in Sect.~\ref{astro_62}, of all the pure PAHs and closely related species studied to date, only those that contain internal N atoms can reproduce the peak position of the class A band, i.e. that which peaks at 6.2 \mum\, \citep{Peeters:prof6:02, Hudgins:05}.  Although Paper 1 and the discussion in Sect.~\ref{astro_33} to Sect.~\ref{astro_112} above make it clear that negatively charged PAHs are important in determining the overall emission spectrum, the spectra of nitrogenated PAH {\it anions} were not previously considered. To explore the effect negative charge has on the spectra of nitrogenated PAHs, we computed the spectra of 3N-coronene (C$_{23}$NH$_{12}$) and 4N- and 5N-circumcoronene (C$_{53}$NH$_{18}$) with single nitrogen atom substitution. The peak position of the strong CC$_{str}$ band corresponding to the cations and anions of these species, as well as their non-nitrogenated counterparts, are listed in Table~\ref{t7}. {\it Table~\ref{t7} shows that the addition of an electron to a nitrogenated PAH redshifts the CC$_{str}$ from class A to class B. This makes it clear charge determines the position of the 6.2-6.3 \mum\, astronomical feature. This is in contrast with the pure large PAHs considered here and in paper I, where anions peak slightly to the blue of the cations. If born out by further study, this implies that the interstellar PAH population could well be comprised largely of nitrogenated PAHs as described in \citet{Hudgins:05}, with the relative intensities of the Class A band to Class B band directly reflecting the PAH cation to anion ratio.} 
A thorough assessment of this suggestion requires knowledge of the IR spectroscopic properties of a wide range of nitrogenated PAHs, information which is currently unavailable. 

\clearpage
\begin{table}
\caption{\label{t7} Computed positions of the dominant 6.2 \mum\, bands for a range of singly substituted nitrogenated PAHs and their parent compounds.}
\begin{center}
\begin{tabular}{lcc}
\hline \\[-5pt]
Species & \multicolumn{2}{c}{Dominant 6.2 \mum} \\
 & \multicolumn{2}{c}{CC$_{str}$ position}\\
 & Cation & Anion \\[5pt]
 Coronene (C$_{24}$H$_{12}$) & 6.442 & 6.418 \\
 3N-coronene\tablenotemark{a} (C$_{23}$NH$_{12}$) & 6.191 & 6.252 \\
 Circumcoronene  (C$_{54}$H$_{18}$) & 6.365 & 6.364 \\
 4N-circumcoronene (C$_{53}$NH$_{18}$) & 6.223 & 6.348 \\
 5N-circumcoronene (C$_{53}$NH$_{18}$) & 6.204 & 6.281\\[5pt]
\hline \\[-5pt]
\end{tabular}
\end{center}
\tablenotetext{a}{see \citet{Hudgins:05} for a description of this nomenclature. The values in this table are computed using higher precision and hence are slightly different from and supersede the previous values.}
\noindent
\end{table}
\clearpage

We now consider the 7.7 \mum\, band. The results presented in earlier sections and the spectra in Figure~\ref{fig_6to9} reinforce the conclusions drawn in Paper 1, namely that the astronomical 7.7 \mum\, band is produced by overlapping bands from a mixture of small and large PAH cations and anions, with ``small" PAHs contributing more to the 7.6 \mum\, component and large PAHs (C$>$$\sim$100) more to the 7.8 \mum\, component. As also pointed out in Paper 1, and reinforced here, this implies that the variation in the peak position of the 7.8 \mum\, component may be related to the variation in the population of large PAH cations and anions since negatively charged large PAHs emit at slightly longer wavelengths than do the PAH cations. In this regard, it is interesting to note the following observational facts. First,  \citet{Peeters:prof6:02} found a general correlation between the peak position of the 6.2 and 7.7 \mum\, PAH bands. Hence, this is consistent with charge determining the peak position of both features.  Second, \citet{Bregman:05} found a dependence of the 7.7 \mum\, centroid with G$_0$/n$_e$, which determines the charge balance, in three reflection nebulae. These authors interpret this by either a charge variation (with the 7.85 \mum\, band originating in anions and the 7.65 \mum\, band originating in PAH cations) or a variation in the physical nature of the emitting PAHs. Third, \citet{Joblin:08} decomposed the mid-IR emission of planetary nebulae and HII regions into six components including a PAH population with a 7.9 \mum\, band that was attributed to very large cations and/or anions on the basis of our Paper 1.

Turning now to the region longward of 8 \mum, Figures~\ref{fig_average}b, ~\ref{fig_average}e, and ~\ref{fig_6to9} show that only large compact PAHs can produce a distinctive band near 8.6 \mum. Reiterating the conclusions drawn in Paper 1, the prominent bands from large compact PAH cations and anions overlap, with the anions contributing more strongly at slightly longer wavelengths. The data presented here and in Paper 1, taken with the correlation between the 6.2, 7.7 and 8.6 \mum\, bands \citep{Peeters:prof6:02}, strongly suggest that the astronomical 7.8 \mum\, feature and the 8.6 \mum\, band originate primarily in compact, large cationic and anionic PAHs, with the specific peak position and profile reflecting the cation to anion ratio in any given object (Fig.~\ref{fig_final}). Hence, this dependence on charge state for the 6.2, 7.7 and 8.6 \mum\, bands is likely responsible for the observed relationship between the peak positions of these three bands \citep{Peeters:prof6:02}.

\clearpage
\clearpage
 
 \paragraph{The 10-15 \mum\, Features} We end with the bands that are carried by PAHs that span a wide size range.  While species with fewer than about 80 C atoms are predicted to contribute some 10 to 15\% to the emission in these features, the bulk of the intensity is expected to be carried by PAHs containing between about 80 to several hundred carbon atoms \citep{Schutte:model:93}. Very large ÔaromaticÕ species with over 500 carbon atoms and extending up to 3000 C atoms could, in principle, carry as much as 20\% of the total emission in this region \citep{Schutte:model:93}. While it is difficult to imagine how such large individual PAHs could be readily formed under typical interstellar conditions, the presence of large PAH-related species such as PAH clusters and simple amorphous carbon
particles within the emission zones is likely \citep{ATB, Rapacioli:06, Rhee:07}. Clearly, the issue of the size range of the PAHs that contribute to the bulk of the emission in each wavelength region warrants reconsideration now that the spectroscopic properties of PAHs are so much better understood.  The earlier work of \citet{Schutte:model:93} was based on ÔgenericÕ PAH properties that were deduced from the astronomical emission spectra before any laboratory or computational PAH spectra were available.  
 
\citet{Hony:oops:01} have shown that the spectra of NGC\,7027 and IRAS\,18317 (Figure~\ref{fig_oops}) represent limiting cases of the 10 to 15 \mum\, astronomical PAH features of their sample. Figures ~\ref{fig_average}c, \ref{fig_average}f, and \ref{fig_oops}, along with Table~\ref{t6}, place strong constraints on the structures of large astronomical PAHs. The large PAHs in objects showing the NGC\,7027 type of spectrum have roughly ten times more solo than duo or trio hydrogens. At the other extreme, objects with the IRAS\,18317 type of spectrum reflect a large PAH population with structures that are not as dominated by solo hydrogens, but still have more solo than duo or trio hydrogens. In both cases the contributions of edge rings that add trio hydrogens to the PAHs is small. Examples of PAH structures with solo/duo and solo/trio hydrogen ratios that produce the NGC\,7027 and IRAS\,18317 type of spectra are shown in Figure~\ref{eg_structure}. Clearly, the largest PAHs in the astronomical PAH population that produce emission features similar to those from NGC\,7027 are dominated by compact structures with very regular edges. However, the structures of the large PAHs implied by the
10 to 15 \mum\, spectra of objects with spectra similar to that of IRAS\,18317 have irregular edges, and therefore more bay regions than deduced from their spectra in the 3.3 \mum\, region. This difference
in deduced edge structures arises because the features in these two regions originate from distinctly different portions of the astronomical PAH population in the same emission zone. The 3.3 \mum\, band is dominated by PAHs containing between about 30 to 70 C atoms whereas the features between10 and 15 \mum\, are produced by PAHs comprised of some 80 to several hundred C atoms.
 
While the spectra in Figure~\ref{fig_oops} show that large neutral PAHs can account for the bulk of the 11.2 \mum\, band, it also shows that these cannot produce the long wavelength wing \citep[e.g.][]{Witteborn:89}.  A number of plausible suggestions have been made to account for the wing, including different PAHs with slightly shifted solo bands, and hot band emission shifted due to anharmonicity \citep{Barker:anharm:87, Pech:prof:01, Verstraete:prof:01}.  Here we add another possible contributor to the wing, emission from negatively charged large PAHs. Figures~\ref{fig_average}c, \ref{fig_average}f and \ref{fig_oops} show that the solo CH$_{oop}$ bands in large PAH anions fall at the correct position and are sufficiently intense to add to the wing. In view of the important role PAH anions play in determining the precise peak positions and profiles of all the other major bands, it is reasonable that they too contribute to the long wavelength bands.   
Thus, if the importance of the anion contribution to the wing can be determined, the 10.5 to 11.5 \mum\, spectrum of regions emitting the PAH features can be used to probe the relative amounts of large PAH cations, anions, and neutral species present since the generally weak 11.0 \mum\, band likely arises from cations \citep{Hudgins:tracesionezedpahs:99}. \\
 
These conclusions, about the size and charge of the PAHs producing the prominent features for the class A and class B emission objects are presented schematically in Fig.~\ref{fig_final}.

\section{Conclusions}
\label{conclusion}

The main conclusions of this work, described in Sect.~\ref{astro_sum},  can be summarized as follows: 
\begin{itemize} 
\item The vast majority of the PAHs which produce the well-known emission spectra are compact structures with regular edges.  Irregular edge structures which produce bay regions or quadruply adjacent H atoms are the exception.

\item All the PAH charge states contribute to the 3.3 \mum\, feature. These are the smallest members of the astronomical PAH population.

\item The majority of astronomical PAHs may well contain nitrogen within their hexagonal network.

\item The 6.2 \mum\, band can only be produced by PAHs that contain nitrogen. The preliminary results presented here suggest that the ratio of the class A to class B components of this feature may be a measure of PAH cation to anion ratio. 

\item  The 7.7 \mum\, complex is comprised of a mixture of small and large PAH cations and anions with the small species contributing to the 7.6 \mum\, component and the large species to the 7.8 \mum\, component. In contrast, the 8.6 \mum\, band arises exclusively from large, compact PAH cations and anions.
Band positions and profiles directly reveal the PAH cation to anion ratio and reflect PAH size in the emission zones. 

\item The 11 to 13 \mum\, features reveal  the structures of the largest PAHs in the emitting population and the relative amounts of PAH neutrals, cations, and anions in that population.

\end{itemize}

\acknowledgements 
We would like to thank Doug Hudgins for fruitful, animated discussions and his insightful advise regarding Fig. 13. We are particularly grateful to an anonymous referee for careful reading of the manuscript. We very gratefully acknowledge sustained support
from NASA's Long Term Space Astrophysics and Astrobiology Programs,
and the Spitzer Space Telescope Archival and General Observer Program.

\end{document}